\documentclass[sigconf]{acmart}

\usepackage{multirow}
\usepackage{makecell}

\definecolor{pink}{RGB}{0,0,0}

\AtBeginDocument{%
  \providecommand\BibTeX{{%
    \normalfont B\kern-0.5em{\scshape i\kern-0.25em b}\kern-0.8em\TeX}}}

\copyrightyear{2024}
\acmYear{2024}
\setcopyright{rightsretained}
\acmConference[CHI '24]{Proceedings of the CHI Conference on Human Factors in Computing Systems}{May 11--16, 2024}{Honolulu, HI, USA}
\acmBooktitle{Proceedings of the CHI Conference on Human Factors in Computing Systems (CHI '24), May 11--16, 2024, Honolulu, HI, USA}
\acmDOI{10.1145/3613904.3642067}
\acmISBN{979-8-4007-0330-0/24/05}

\begin{document}

\title{QuadStretcher: A Forearm-Worn Skin Stretch Display for Bare-Hand Interaction in AR/VR}

\author{Taejun Kim}
\affiliation{%
  \institution{HCI Lab, KAIST}
  \city{Daejeon}
  \country{South Korea}}
\email{taejun.kim@kaist.ac.kr}

\author{Youngbo Aram Shim}
\affiliation{%
	\institution{HCI Lab, KAIST}
	\city{Daejeon}
	\country{South Korea}}
\email{youngbo.shim@kaist.ac.kr}

\author{Youngin Kim}
\affiliation{%
	\institution{HCI Lab, KAIST}
	\city{Daejeon}
	\country{South Korea}}
\email{youngin3737@gm.gist.ac.kr}

\author{Sunbum Kim}
\affiliation{%
	\institution{HCI Lab, KAIST}
	\city{Daejeon}
	\country{South Korea}}
\email{ksb4587@kaist.ac.kr}

\author{Jaeyeon Lee}
\affiliation{%
	\institution{TACT Lab, UNIST}
	\city{Ulsan}
	\country{South Korea}}
\email{jaeyeonlee@unist.ac.kr}

\author{Geehyuk Lee}
\affiliation{%
  \institution{HCI Lab, KAIST}
  \city{Daejeon}
  \country{South Korea}}
\email{geehyuk@gmail.com}

\renewcommand{\shortauthors}{Kim et al.}

\begin{abstract}
	
The paradigm of bare-hand interaction has become increasingly prevalent in Augmented Reality (AR) and Virtual Reality (VR) environments, propelled by advancements in hand tracking technology. However, a significant challenge arises in delivering haptic feedback to users' hands, due to the necessity for the hands to remain bare. In response to this challenge, recent research has proposed an indirect solution of providing haptic feedback to the forearm. In this work, we present QuadStretcher, a skin stretch display featuring four independently controlled stretching units surrounding the forearm. While achieving rich haptic expression, our device also eliminates the need for a grounding base on the forearm by using a pair of counteracting tactors, thereby reducing bulkiness. To assess the effectiveness of QuadStretcher in facilitating immersive bare-hand experiences, we conducted a comparative user evaluation (n = 20) with a baseline solution, Squeezer. The results confirmed that QuadStretcher outperformed Squeezer in terms of expressing force direction and heightening the sense of realism, particularly in 3-DoF VR interactions such as pulling a rubber band, hooking a fishing rod, and swinging a tennis racket. We further discuss the design insights gained from qualitative user interviews, presenting key takeaways for future forearm-haptic systems aimed at advancing AR/VR bare-hand experiences. 

\end{abstract}

\begin{teaserfigure}
	\centering
	\includegraphics[width=0.75\textwidth]{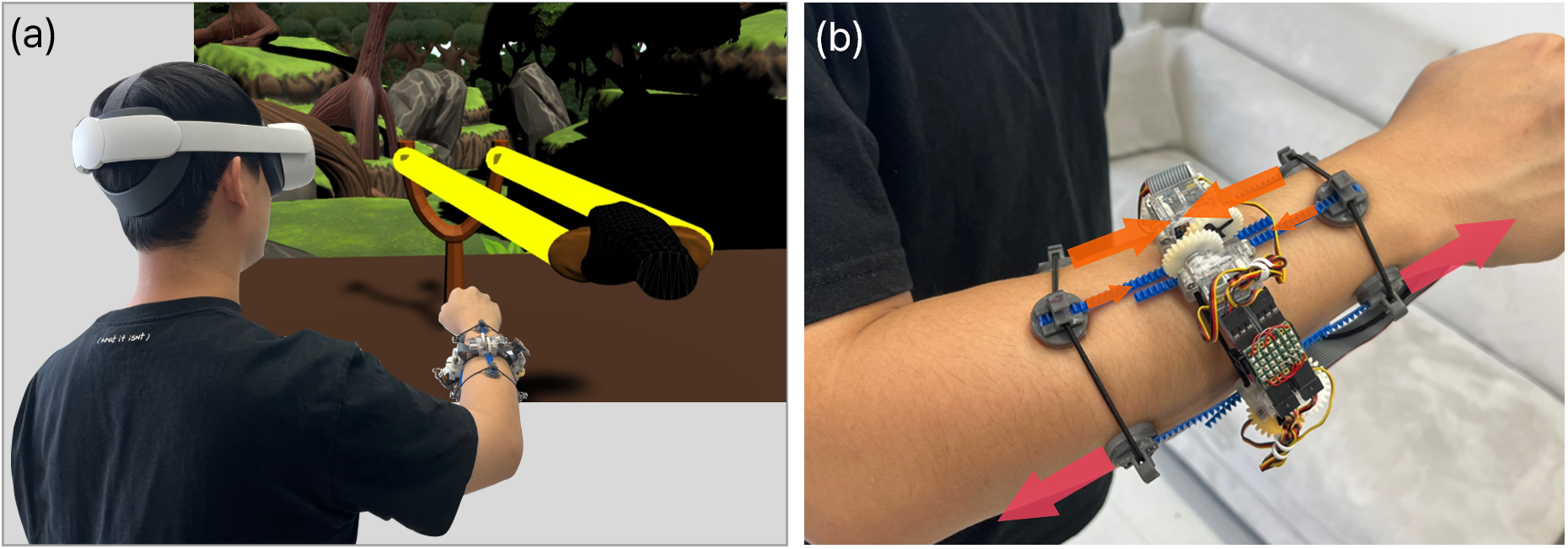}
	\caption{QuadStretcher is a skin stretch display that enhances immersive AR/VR bare-hand experiences. (a) A user is pulling a rubber band in VR, and (b) QuadStretcher renders real-time haptic feedback based on the direction and length of the pulled rubber band.}
	\label{fig:teaser}
	\Description{Left image: a person is wearing a VR headset and pulling a slingshot while wearing the skin stretch device, QuadStretcher. Right image: the closer picture of the QuadStretcher device worn on the forearm. Each pair of counteracting tactors are expanding or contracting the skin.}
\end{teaserfigure}



\maketitle

\section{Introduction}

With advancements in hand tracking technology, major Augmented Reality (AR) and Virtual Reality (VR) headsets now offer seamless bare-hand interactions. For example, Xreal AR glasses~\cite{xrealair}, Microsoft Hololens 2~\cite{microsoftHololens2}, HTC Vive~\cite{htcVivePro2}, Apple Vision Pro~\cite{appleVisionPro}, and Meta Quest series~\cite{metaQuest} offer real-time tracking of finger, palm, and wrist joints. This capability enables users to leverage their intricate hand dexterity and expressiveness in diverse activities. The paradigm of bare-hand interaction is expected to be more prevalent in future AR/VR applications, where social interaction will hold a major theme.

While the future of this paradigm is promising, a significant challenge arises in providing haptic feedback to the user's hands. Previous studies have explored various approaches to provide haptic feedback to the hands for an immersive AR/VR experience, such as using grounded haptic devices~\cite{grange2001overview, massie1994phantom, steed2020docking, sledd2006performance}, body-worn haptic devices~\cite{hirose2001hapticgear, nagai2015wearable}, hand-worn haptic devices~\cite{hinchet2018dextres,perret2018touching}, and handheld devices~\cite{choi2017grabity, choi2016wolverine, choi2018claw, sinclair2019capstancrunch,sun2019pacapa,park2023visuo}. However, these direct feedback methods are not applicable to bare-hand interactions, where the hands must remain unrestricted.

In response to this challenge, recent research has proposed a promising solution of employing haptic feedback on the wrist/forearm area as an indirect approach. The pivotal study by Pezent et al.~\cite{pezent2019tasbi} suggested providing squeeze stimuli to the wrist, with the aim of delivering ``substituted'' feedback when the bare hands are interacting with virtual objects. Using Tasbi~\cite{pezent2019tasbi, pezent2022explorations, pezent2022design}, a wristband that offers precise squeezing control and vibrotactile cues, the researchers tested various AR/VR interactions such as pushing buttons, twisting knobs, and pulling handles. This ``substitutive'' haptic sensation integrated with visuals was reported to significantly enhance the immersion level in AR/VR bare-hand interactions. However, as noted by the authors in their paper~\cite{pezent2022design}, the squeeze expression of Tasbi was essentially a 1-Degree of Freedom (DoF) either in the squeezing or releasing direction. The authors noted that the challenge lies in expressing dynamics associated with forces involving higher DoF, such as the inertia exhibited by a tennis racket.

In this work, we propose QuadStretcher, a skin stretch display featuring four stretching units surrounding the forearm (Figure \ref{fig:teaser}). With the independent control of stretching units on each dorsal, right, ventral, and left side of the forearm, QuadStretcher seizes the potential for rendering dynamic forces with higher DoF. Our device also forgoes the need for a grounding base on the forearm using a pair of counteracting tactors, thereby minimizing the bulkiness typically linked with a firm grounding structure. Additionally, the stretching unit is designed to contract or expand the skin in the longitudinal axis of the forearm, mirroring the biomechanics of forearm muscle tendons \cite{nathan1992isometric} during real-world hand activities.

In a user study with 20 participants, we performed a comparative evaluation of QuadStretcher and the baseline solution, Squeezer, to assess their effectiveness in facilitating immersive bare-hand activities in VR environments. The results showed that QuadStretcher provided a superior experience compared to Squeezer in terms of enhanced expressive power in force direction and a heightened sense of realism, particularly in 3-DoF interactions, such as pulling rubber a band, hooking a fishing rod, and swinging a tennis racket. 

The main contributions of this study are as follows. 

\begin{itemize}
	\item We present a novel artifact, QuadStretcher, a forearm-worn skin stretch display facilitating immersive AR/VR bare-hand experiences.
	\item Through an empirical user evaluation, we revealed that our proposed solution offered a greater expressive capability than the baseline solution, particularly in the dynamics of 3-DoF interactions.
	\item We share design insights and key takeaways for the future forearm-haptic systems, derived from think-aloud interviews conducted during the user study.
\end{itemize}

\section{Related Work}

We first review the paradigm of bare-hand interactions in recent AR/VR studies. Subsequently, we scrutinize wrist/forearm haptic solutions proposed for bare-hand interactions. Lastly, we review the distinctive attributes of skin stretch stimuli.

\subsection{Bare-Hand Interaction in AR/VR}

The paradigm of bare-hand interaction has been increasingly common in AR/VR applications. In AR applications like remote work, health care, and education, bare-hand interaction has been preferred due to the impracticality of performing other tasks with the encumbered hands and carrying additional devices like handheld controllers. In the VR domain, although the dedicated handheld controller was common in its early days, most consumer VR headsets \cite{metaQuest,htcVivePro2,picoVR} now offer precise vision-based hand tracking. This has resulted in a paradigm shift towards unencumbered, bare-hand VR interactions.

The utilization of mid-air hand pointing and finger pinch for selection is now widespread in commercial products \cite{shi2023exploration, pfeuffer2017gaze, metaQuest}. Furthermore, ongoing research is delving into a wide array of interaction techniques using bare hands, including the use of hand gestures for tasks like creative authoring~\cite{arora2019magicalhands}, gaming~\cite{rautaray2011interaction}, or text entry~\cite{nooruddin2020hgr, kim2023star}. A recent study~\cite{pei2022hand} has also put forth an intriguing idea of using a hand as a physical proxy of virtual objects, such as using a thumbs-up posed hand as a joystick. As hand tracking technology for AR/VR headsets continues to advance, bare-hand interaction is expected to become more enriched and prevalent over time. In this context, since many existing haptic feedback methods are not compatible with this trend, it is imperative to establish a haptic solution to let the hands remain bare. 

\subsection{Wrist/Forearm Haptic Solutions for Hand Interaction}

\subsubsection{Prosthetics Studies}

The original idea of the ``substitutive'' haptic feedback on the wrist/forearm area during hand interaction stems from prosthetics studies. A series of research suggested providing haptic stimuli on the residual limb in response to the prosthetic hand actions, aiming to enhance the sense of embodiment~\cite{rohland1975sensory, shannon1979myoelectrically, cipriani2008shared, pylatiuk2006design, saunders2011role, shannon1976comparison}. While early studies have mostly employed electrical stimuli~\cite{shehata2020mechanotactile, rohland1975sensory, shannon1979myoelectrically, cipriani2008shared}, pressure sensation~\cite{marasco2011robotic, antfolk2012sensory}, and vibrotactile stimuli~\cite{shannon1976comparison, saunders2011role, cipriani2008shared}, employing skin stretching has recently gained attention \cite{battaglia2019skin, colella2019novel,stephens2018applying,wheeler2010investigation} for its expressive power in proprioception \cite{battaglia2019skin} and force~\cite{stephens2018applying}.

\subsubsection{Studies in AR/VR domain}

In line with the bare-hand interaction paradigm, several studies have explored the wrist/forearm haptic solutions to support freehand activities. Moriyama et al.~\cite{moriyama2018development} developed a forearm-worn haptic device engineered to provide pressure stimuli through tactors moving in directions perpendicular to the arm, as well as skin stretch stimuli through tactors moving in directions parallel to the arm's skin. They reported an improvement in subjective realism ratings when users employed their device while gripping and lifting an object in a VR environment. Sarac et al.~\cite{sarac2022perceived} examined a haptic bracelet with two tactors positioned on the dorsal and ventral sides of the forearm. The bracelet had the capability to provide either normal or shear force feedback, depending on its grounded direction. Users were tasked with discerning the stiffness of virtual objects with hand exploration, while the bracelet was rendering the feedback. They showed that haptic feedback on the forearm, both in normal and shear direction, facilitated an accurate perception of stiffness. However, these solutions were confined to investigating haptic feedback just at specific points on the arm, rather than encompassing the surrounding area of the wrist/forearm. This limited the capability of haptic devices to convey enriched sensory experiences.

\begin{figure}[t]
	\centering
	\includegraphics[width=8.4cm]{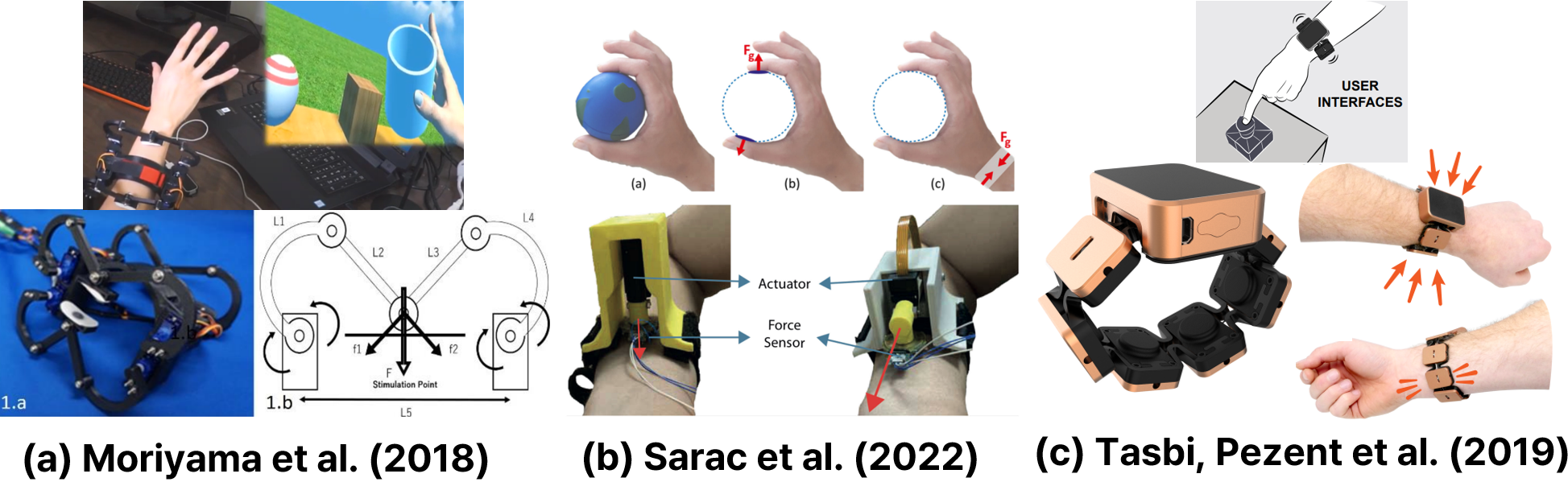}
	\caption{Previously proposed wrist/forearm haptic solutions to support bare-hand activities. (a) Moriyama et al. (2018)~\cite{moriyama2018development}, (b) Sarac et al. (2022)~\cite{sarac2022perceived}, and (c) Pezant et al. (2019)~\cite{pezent2019tasbi}}
	\label{fig:relatedWorkForearmHaptic}
	\Description{Left image: the forearm-worn haptic device proposed by Moriyama et al. (2018). Middle image: the forearm-worn haptic device proposed by Sarac et al. (2022). Right image: the wrist-worn squeezing device named Tasbi, proposed by Pezent et al. (2019)}
\end{figure}

Tasbi~\cite{pezent2019tasbi, pezent2022explorations, pezent2022design} is a wristband that provides precise control of squeezing surrounding the wrist (Figure \ref{fig:relatedWorkForearmHaptic}c), along with vibration cues. The use of Tasbi was extensively explored in various VR hand activities, such as pushing buttons, rotating knobs, pulling handles, rendering object weights and inertia \cite{pezent2022explorations}. Through multiple user studies, Tasbi demonstrated its ability to significantly enhance the level of immersion during virtual hand interactions.

However, as the authors noted in the paper \cite{pezent2022explorations}, the wrist squeeze was essentially a 1-DoF modality either in squeezing or releasing directions, thereby showing a limited expressive power especially for dynamics with higher DoF forces. In this study, we propose QuadStretcher which is capable of delivering enhanced sensory experiences with four independently controlled stretching units. The design and implementation of QuadStretcher is described in the following Section 3. (We note that this device has previously been demonstrated in ACM SIGGRAPH 2022~\cite{shim2022quadstretchSiggraph} and CHI 2022~\cite{shim2022quadstretchChi}).

\subsection{Skin Stretch Stimulus}

Skin stretch is a skin deformation stimulation that occurs when a tactor moves the skin tangentially. It is also referred to as lateral or tangential displacement, or shear feedback in the literature. The skin stretch is sensed by a mechanoreceptor known as the Ruffini ending, and the stimulation is transmitted to the central nervous system through the SAII afferent~\cite{johnson2001roles,jones2006human,mcglone2010cutaneous}. The unique sensation of skin stretch, distinct from other types of stimuli such as vibrations~\cite{tan2020methodology,kim2021heterogeneous,youn2022wristmenu,kim2023human,park2022vibration,park2020augmenting,park2019realistic}, has been extensively explored with numerous haptic devices for VR and wearable computing~\cite{caswell2012design,provancher2014creating,yamashita2018gum,pacchierotti2017wearable,minamizawa2007gravity,quek2013sensory,quek2014augmentation,leonardis2015wearable,schorr2017fingertip,ion2015skin,wang2019masque,wang2020gaiters}. 

Our idea of incorporating skin stretch stimuli on the forearm for bare-hand interaction was motivated by the biomechanics of forearm muscle tendons, which contract and expand while hands are interacting with physical objects~\cite{corcondilas1964effect,nathan1992isometric}. We conduct a comparative user assessment of our skin stretch display with the baseline solution, Squeezer, and reveal its effectiveness.

\section{QuadStretcher: A Forearm-Worn Skin Stretch Display}

\begin{figure}[t]
	\centering
	\includegraphics[width=7cm]{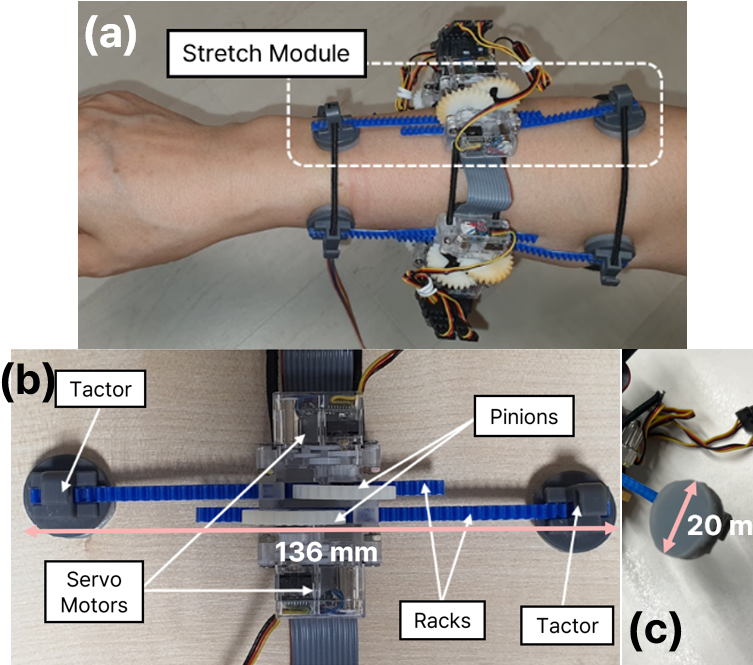}
	\caption{(a) The QuadStretcher worn on the forearm. (b) The top view of the stretch module. (c) The diameter of the stretch tactor.}
	\label{fig:quadStretchDevice}
	\Description{Top image: the picture of the QuadStretcher device worn on the forearm. One stretch module is highlighted with a dotted rectangle. Bottom left image: the top view of the stretch module which is not worn on the human arm. Components are tactor, pinions, servo motors, and racks. The width is 136 mm. Bottom right image: the diameter of the stretch tactor is 20 mm.}
\end{figure}

In this section, we provide a comprehensive description of QuadStretcher, covering its design process, final implementation, range of skin displacement, range of comfortable stretch output, psychophysical investigation of Just Noticeable Differences (JNDs), and confusion matrices of possible stretch stimuli set.

\subsection{Design Process}

We adopted an iterative approach, focusing on enhancing expressiveness, wearability, comfort, and reducing bulkiness through the refinement of stretch direction, grounding methods, and the use of suitable actuators.

Our initial design decision was to determine the direction for skin stretching. Among choices between longitudinal and lateral directions with respect to the forearm axis, we opted for longitudinal skin stretching for two reasons. Firstly, previous research has shown that longitudinal stretching is more easily recognized compared to lateral stretching when the same degree of stretch is applied~\cite{stephens2018applying}. Secondly, we expected that users would associate better perceptual mapping from the forces applied to hands to the longitudinal stretch, compared to the lateral stretch due to the biomechanics of forearm muscle tendons, which contract or expand in the longitudinal direction~\cite{nathan1992isometric}. While this suggested mental model may not precisely mirror anatomical reality, we anticipated that this design would help users form a more intuitive connection between the stimuli and the virtual force applied to their hands.

Another pivotal decision was to choose not to employ a base frame, which is a common practice in the design of stretch actuators for firm grounding~\cite{caswell2012design,moriyama2018development,wang2020gaiters,yamashita2018gum}. This design decision was motivated by the drawbacks associated with the use of the base frame for each tactor, making the whole device bulky and less wearable. Instead, we devised an alternative solution of utilizing a pair of counteracting tactors along the forearm (Figure \ref{fig:quadStretchDevice}a). This design also averted the compression of the skin around the stretching area by the base frame, which might degrade the stretch perception \cite{gleeson2010improved}.

Towards better wearability, we adopted small and light actuators and flexible structural elements. The final prototype weighed 147 g.

\subsection{Implementation}

The device consists of four stretch units, as shown in Figure \ref{fig:quadStretchDevice}a. Each of the stretch units has two counteracting tactors, and each tactor was driven by a servo motor (HiTec HS-40) via a rack-and-pinion mechanism. The two servo motors move synchronously and drive the two tactors in opposite directions. The racks are 3 mm wide, 95 mm long, and made of a flexible plastic material (DR0.8-2000, KHK). The tactors are silicone discs with a diameter of 20 mm and a thickness of 3 mm and are attached to the end of the rack gears using 3D-printed parts. The four stretch units are mounted around the forearm with elastic rubber strings and locks.

\begin{figure}[t]
	\centering
	\includegraphics[width=7.5cm]{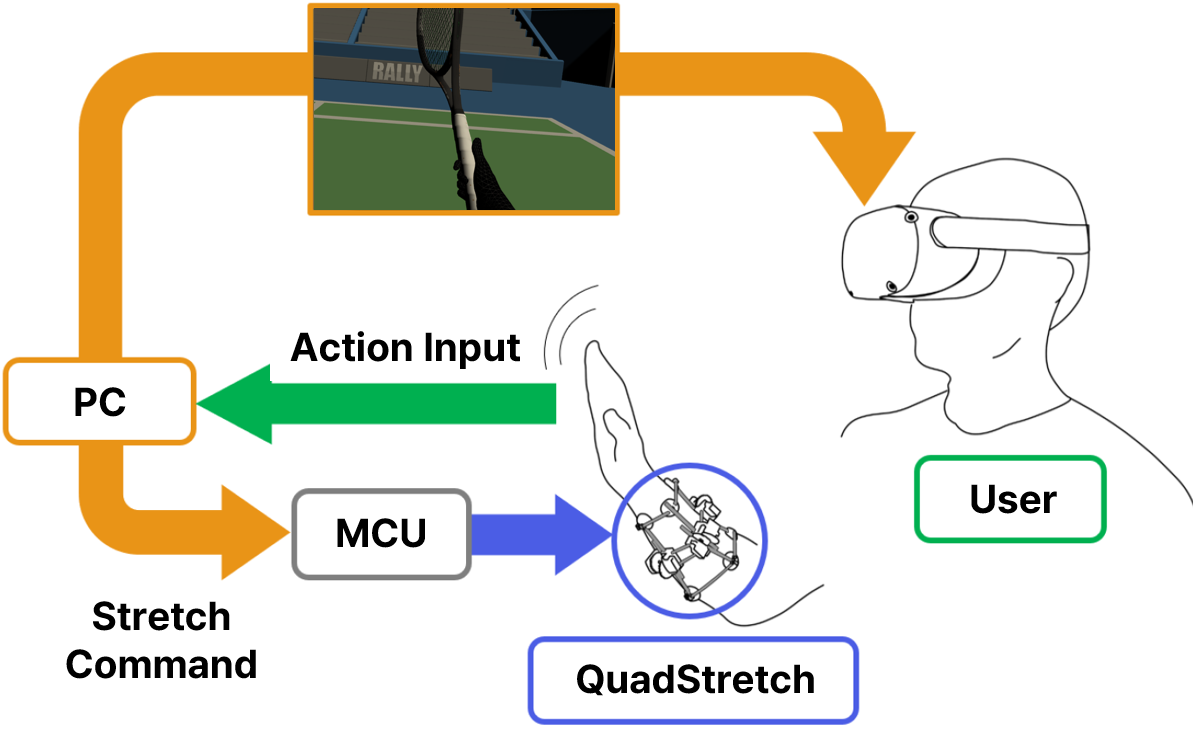}
	\caption{The system pipeline of using QuadStretcher.}
	\label{fig:quadStretchSystemPipeline}
	\Description{The illustration of the system pipeline of the QuadStretcher system. User looks at VR scenes computed from the PC. When a user makes an input while wearing the QuadStretcher, PC calculates the corresponding control signal of the stretch and delivers it to the MCU. The user then get real-time haptic feedback from the QuadStretcher in response to their action.}
\end{figure}

Corresponding to the user hand actions in VR applications run by Unity, an Arduino (AVR) board received commands from the Unity application on a PC (Figure \ref{fig:quadStretchSystemPipeline}). The Arduino board then relayed the commands to the four stretch modules via a PWM driver board (SunFounder, PCA9685 16-Channel 12-Bit PWM servo driver). The servo motors were powered at 6V. 

Each tactor under no load (i.e., when the tactor is not worn on the arm) moves approximately by $\pm11.0$ mm from the neutral position when the stretch module expands or contracts maximally. This means that the distance between the two counteracting tactors can maximally increase or decrease by 22 mm. In the remainder of this paper, the control signal for each stretch unit is the ``tactor displacement under no load (mm)''. The control signals of $-11.0$, 0, and 11.0 mm correspond to the maximum contraction, neutral, and maximum expansion positions of the tactors, respectively (Figure \ref{fig:skinDisplacementResult}a). According to the specifications of the servo motors and the pinion gears, the maximum tactor speed and force are 206 mm/s and 6.8 N, respectively.

\subsection{Range of Skin Displacement}
\label{rangeOfSkinDisplacementSection}

\begin{figure}[t]
	\centering
	\includegraphics[width=8.4cm]{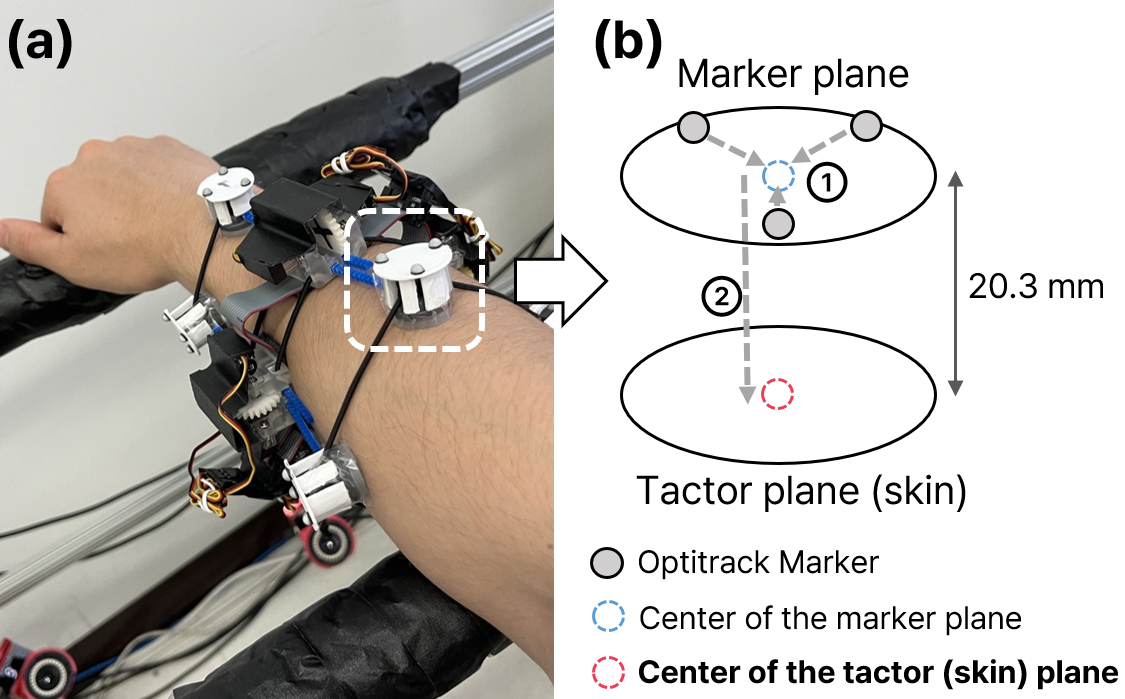}
	\caption{(a) The environment for the skin displacement measurements. (b) The process of computing the center of the tactor's bottom plane in contact with the skin: calculating the center of the marker plane first, and translating it by 20.3 mm. This was done while ensuring the tactor plane was parallel to the marker plane with a 3D-printed rigid add-on structure.}
	\label{fig:skinDisplacementTestSetup}
	\Description{Left image: an arm with the QuadStretcher worn is lying on the Optitrack platform for the skin displacement measurement. Right image: the illustration of how to compute the center of the tactor's bottom plane.}
\end{figure}

\begin{figure}[t]
	\centering
	\includegraphics[width=8.4cm]{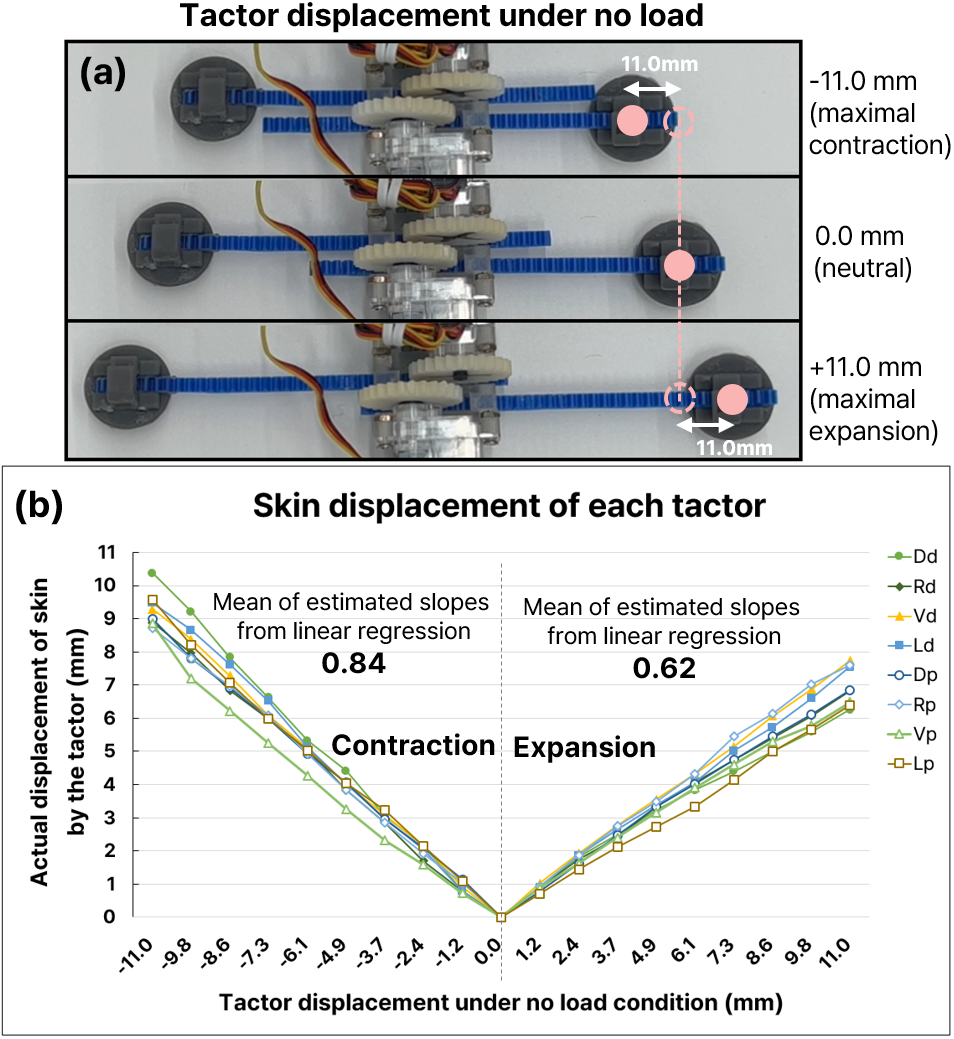}
	\caption{(a) The amount of tactor displacement under no load: $-11.0$, 0.0, and +11.0 mm shown by top, middle, and bottom images, respectively. (b) The measured correlation between the control signal, the tactor displacement under no load condition (mm), and the amount of actual skin displacement (mm). D, R, V, and L denote the dorsal, right, ventral, and left side of the forearm, respectively, while the suffix 'd' and 'p' denote the distal and proximal sides. For example, 'Dd' denotes the tactor in the dorsal forearm at a more distal location.}
	\label{fig:skinDisplacementResult}
	\Description{Top image: the three QuadStretcher images, with a maximal contraction, neutral, and maximal expansion states. Bottom image: the plot of the actual displacement of skin by the tactor (mm) over the tactor displacement under no load condition (mm)}
\end{figure}

The range of actual skin displacement can be smaller than the control signal, ``tactor displacement under no load'', because the output torque of the servo motors would encounter resistance from the skin. Consequently, the flexible racks connected to the tactor can bend due to the grounding of the tactor on the skin. To reveal the skin displacement range that QuadStretcher offers, we conducted a measurement study with 6 participants (2 females, 4 males; age: M = 25.0 years, SD = 2.8 years).

We used an optical tracker, OptiTrack with 11 cameras, for measuring skin displacement. Since it was not feasible to directly attach a marker to the skin which was in contact with the tactor, we attached three markers on the plane parallel to the tactor with 3D-printed add-on structures. We then computed the center of the marker plane and translated it by 20.3 mm to derive the center of the tactor's bottom plane in contact with the skin (Figure \ref{fig:skinDisplacementTestSetup}b). Finally, we measured the displacement of that virtual point on the skin while driving the QuadStretcher. The tactor plane was ensured to be parallel to the marker plane with rigid structures. Participants wore the device in its neutral position on their right forearm, and rested the arm on the Optitrack platform as shown in Figure \ref{fig:skinDisplacementTestSetup}a. We used a medical, double-sided, adhesive tape between the tactor and the skin to avoid any slippage. (The tape was used only for this measurement, not in the later studies) Expansion commands from 0 mm to +11.0 mm and contraction commands from 0 mm to $-11.0$ mm in the step of 10 were provided to the device, and the amount of skin displacements was recorded. This measurement was repeated for the four stretching units.

Figure \ref{fig:skinDisplacementResult}b illustrates the results. Through linear regression analysis, we determined the estimated ratio between the control signal and the mean skin displacement. For contraction, this ratio ranged from 0.78 (Vp) to 0.95 (Dd), with a mean ratio of 0.84. For expansion, the ratio ranged from 0.55 (Dd) to 0.70 (Rp), with a mean of 0.62. The amount of skin displacement showed a significant difference between contraction and expansion (\textit{p} < .001, \textit{F}(1,5) = 254.455, one-way RM ANOVA). We want to note that there was no tactor slippage observed, as the tactors were firmly affixed to the skin using medical adhesive tapes. The asymmetric stretching of the skin occurred because the flexible rack underwent bending during expansion due to an increase in the tactor-to-tactor distance (Figure \ref{fig:skinDisplacementResult}a, bottom image), whereas the rack remained unbent during contraction. We further discuss this circumstance in later Section \ref{futureWorkAndLimitation}. 

\subsection{Comfortable Output Range}

Using the full output range of the stretch stimuli may incur user discomfort, so we explored the comfortable output range with 9 participants (3 females, 6 males; age: M = 24.1 years, SD = 2.8 years). Participants wore the device on their right forearm and were instructed to increment the command signal by either 10\% (1.1 mm) or 1\% (0.11 mm) with keyboard buttons. They reported the value at which they began to experience any discomfort. If no discomfort was reported until the end, the maximum output (11.0 mm) was recorded. This procedure was repeated for each of the four sides (dorsal, right, ventral, and left) of the stretching units, for both contraction and expansion (4 sides $\times$ 2 stretch types).

While there were individual differences in the reported comfortable limits, the median value for all conditions was consistently 11.0 mm (i.e., the maximum range). With the exception of two participants who exhibited unusually small comfortable limits, all participants reported values larger than 8.6 mm across all conditions. Based on this analysis, we determined to confine the control signals within the $\pm$8.6 mm range for the remainder of this study. 

\subsection{Just Noticeable Differences}

\begin{table}[t]
	\caption{The reported JNDs \color{pink}(Weber fraction \textit{k}, in percentage) \color{black}for each side of the forearm and stimulus type. The control signal of the device (i.e., the tactor displacement under no load) served as the stimulus variable for these reported JNDs. For the correlation between the amount of the actual skin displacement and the control signal, refer to Figure \ref{fig:skinDisplacementResult}.}
	\centering
	\resizebox{8.4cm}{!}{%
		\begin{tabular}{@{}c@{\hspace{4pt}}c@{\hspace{4pt}}c@{\hspace{4pt}}c@{\hspace{4pt}}c@{}}
			\hline
			\textbf{} &
			\textbf{Dorsal} &
			\textbf{Right} &
			\textbf{Ventral} &
			\textbf{Left} \\ \hline
			\textbf{Expansion} &
			\begin{tabular}[c]{@{}c@{}}1.3 mm (30.2\%)\\ SD: 0.5 mm\end{tabular} &
			\begin{tabular}[c]{@{}c@{}}1.4 mm (32.6\%)\\ SD: 0.4 mm\end{tabular} &
			\begin{tabular}[c]{@{}c@{}}1.5 mm (34.9\%)\\ SD: 0.4 mm\end{tabular} &
			\begin{tabular}[c]{@{}c@{}}1.5 mm (34.9\%)\\ SD: 0.4 mm\end{tabular} \\ \hline
			\textbf{Contraction} &
			\begin{tabular}[c]{@{}c@{}}1.2 mm (27.9\%)\\ SD: 0.4 mm\end{tabular} &
			\begin{tabular}[c]{@{}c@{}}1.4 mm (32.6\%)\\ SD: 0.3 mm\end{tabular} &
			\begin{tabular}[c]{@{}c@{}}1.2 mm (27.9\%)\\ SD: 0.4 mm\end{tabular} &
			\begin{tabular}[c]{@{}c@{}}1.5 mm (34.9\%)\\ SD: 0.4 mm\end{tabular} \\ \hline
		\end{tabular}%
	}
\end{table}

The Just Noticeable Difference (JND) is the minimum detectable change in the intensity of a stimulus ~\cite{jones2012application}. Understanding JNDs can help designers know usable distinct levels of haptic feedback to be employed in AR/VR applications. It is important to note that, in this JND measurement, the stimulus variable was the control signal of the device, not the actual skin displacement, as the objective of this study is to comprehend the device, QuadStretcher. For those interested in the JND of skin displacement, refer to the relationship between the two, shown in Figure \ref{fig:skinDisplacementResult}b.

Participants (8 participants, 3 females, 5 males; age: M = 24.3 years, SD = 3.0 years) wore the device on their right forearm, and put on noise-canceling headphones to block the motor sound from the device. A JND measurement session employed the 2-Down-1-Up Staircase method~\cite{wang2020gaiters}. Each trial presented three stretch stimuli to participants, with two at a reference level and the other differing from the reference by $\Delta S$. Participants engaged in a forced-choice selection among the three stimuli that they felt were different (3-AFC). The reference level was set at 4.3 mm for expansion or $-4.3$ mm for contraction, which was halfway to the comfortable output limit. Initiated at 1.7 mm (40\% of the reference level), the value of $\Delta S$ was adjusted incrementally or decrementally by 20\% for the initial three reversals and by 4\% for the subsequent five reversals. The JND was determined by averaging $\Delta S$ across the last four reversals. JND estimation sessions were repeated for the 8 conditions (4 sides $\times$ 2 stretch types), with the order of conditions counterbalanced across participants using a balanced Latin square. 

Table 1 presents the measured JNDs for all conditions. A two-way RM ANOVA revealed no significant difference in the reported JNDs across different stimulating positions and stretch types. The average JND of the control signal was 1.4 mm (Weber fraction of 33.2\%) for expansion and 1.3 mm (30.8\%) for contraction. These results imply that the control signal should be about 30-33\% higher or lower than the former value for users to perceive a difference. While the reported JNDs were slightly larger than those reported in other body locations such as the face (\textasciitilde25\%~\cite{wang2019masque}) or the fingertip (\textasciitilde20\%~\cite{provancher2009fingerpad}), our device still ensures a sufficient dynamic range for the haptic stimuli, offering multiple distinguishable sensations.

\subsection{Stimuli Discrimination Test: Confusion Matrix}

We conducted a stimuli discrimination test to evaluate how well users could discriminate the positions, i.e., dorsal, right, ventral, or left, and stretch types, i.e., contraction or expansion, of the haptic stimuli generated by QuadStretcher (8 participants, 2 females, 6 males; age: M = 24.4 years, SD = 3.0 years). Participants wore the device on their right forearm and used their left hand to respond via a keyboard. They wore noise-canceling headphones to block the device's sound. During a tutorial session, participants freely explored 8 different stimuli until they became familiar with the testing program. In the subsequent four main sessions, participants were tasked with identifying a randomly presented stimulus among a set of eight stimuli (4 sides $\times$ 2 stretch types), four stimuli (4 sides, under contraction), four stimuli (4 sides, under expansion), and two stimuli (contraction or expansion, all stretching units), respectively. Each stimulus was presented 10 times in each session.

\begin{figure}[t]
	\centering
	\includegraphics[width=8.4cm]{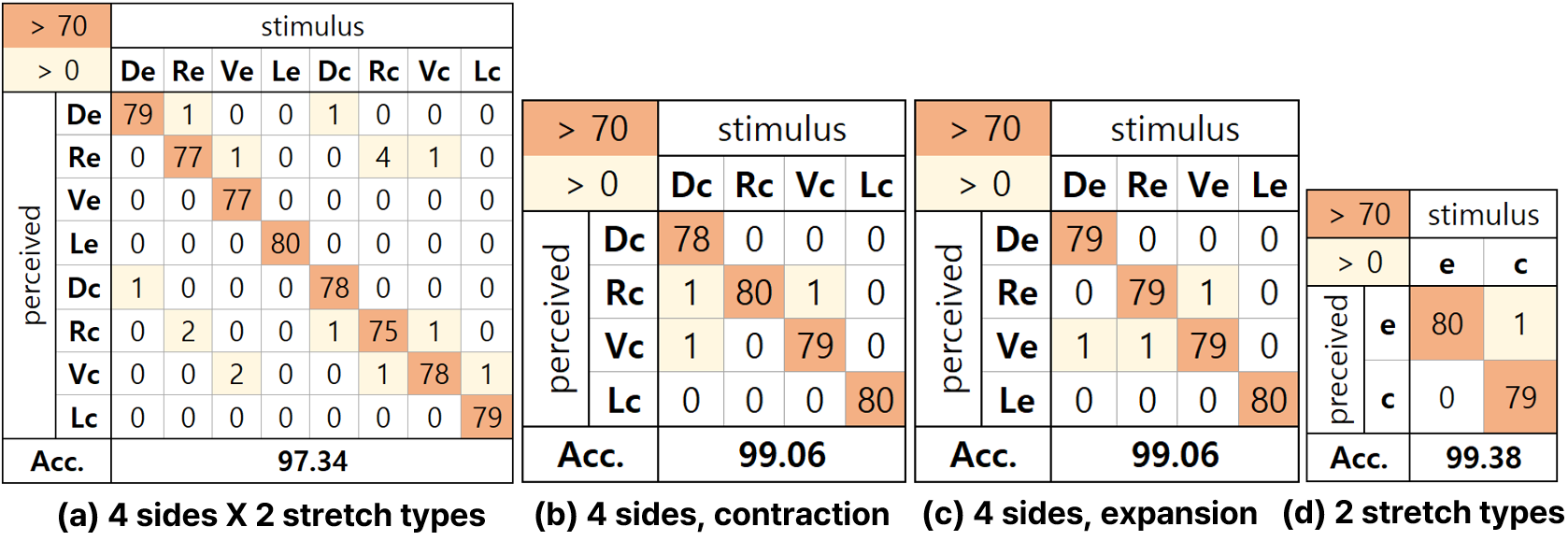}
	\caption{The stimulus-response confusion matrices (a) between 8 stimuli (4 sides $\times$ 2 stretch types), (b) 4 stimuli (4 sides, all contraction), (c) 4 stimuli (4 sides, all expansion), and (d) 2 stimuli (expansion and contraction, all sides). D, R, V, and L denote the dorsal, right, ventral, and left side of the stretching unit, respectively, while the suffixes 'e' and 'c' denote expansion and contraction. For example, 'De' denotes the expansion of the dorsal stretching unit.}
	\label{fig:confusionMatrices}
	\Description{Four confusion matrices. Leftmost: 4 sides by 2 stretch types, eight by eight confusion matrix. One after leftmost: 4sides, contraction. Third one: 4 sides, expansion, Last one: 2 stretch types.}
\end{figure}

Figure \ref{fig:confusionMatrices} displays the confusion matrices for the four sessions with accuracies of 97.34\%, 99.06\%, 99.06\%, and 99.38\%. The result indicates that users were able to easily distinguish all haptic stimuli generated by QuadStretcher.

\section{Implementation of Squeezer}

\begin{figure}[t]
	\centering
	\includegraphics[width=8.4cm]{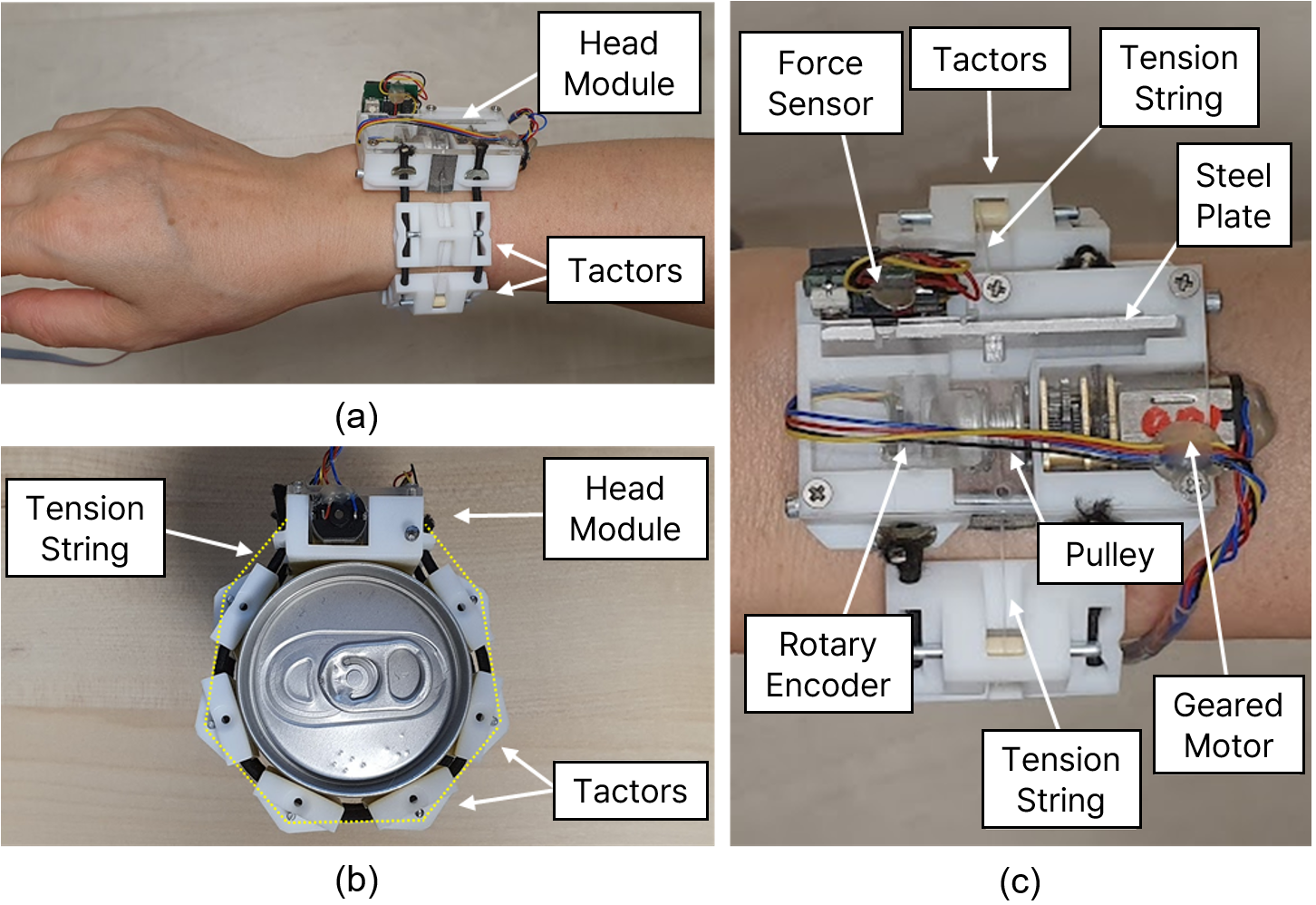}
	\caption{(a) Squeezer worn on a forearm. (b) A squeezer worn on a can showing the head module driving the six tactors using a tension string. (c) The structure of the head module.}
	\label{fig:squeezerDevice}
	\Description{The pictures of the Squeezer device. Top left image: the Squeezer is worn on a forearm. Top bottom image: the top view of the Squeezer worn on a drink can. Right image: the detailed structure of the Squeezer's head module, consists of force sensor, tactors, tension string, steel plate, pulley, geared motor, and rotary encoder.}
\end{figure}

\begin{figure}[t]
	\centering
	\includegraphics[width=6cm]{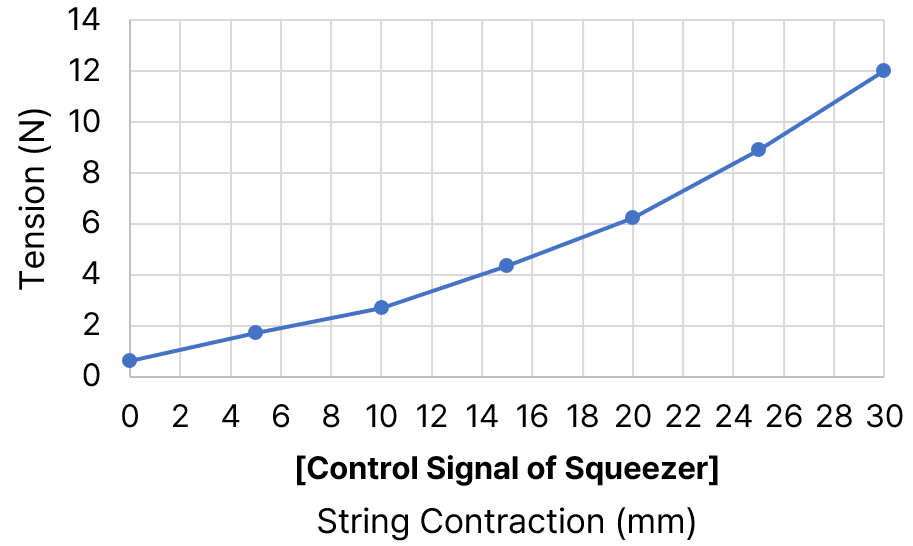}
	\caption{The relationship between the extent of the string contraction and the associated tension exerted around a wrist, as measured by the internal force sensor. }
	\label{fig:squeezerProfile}
	\Description{The point plot of the tension over string contraction, indicating the relationship between the extent of the string contraction and the associated tension.}
\end{figure}

\begin{figure*}[t]
	\centering
	\includegraphics[width=10cm]{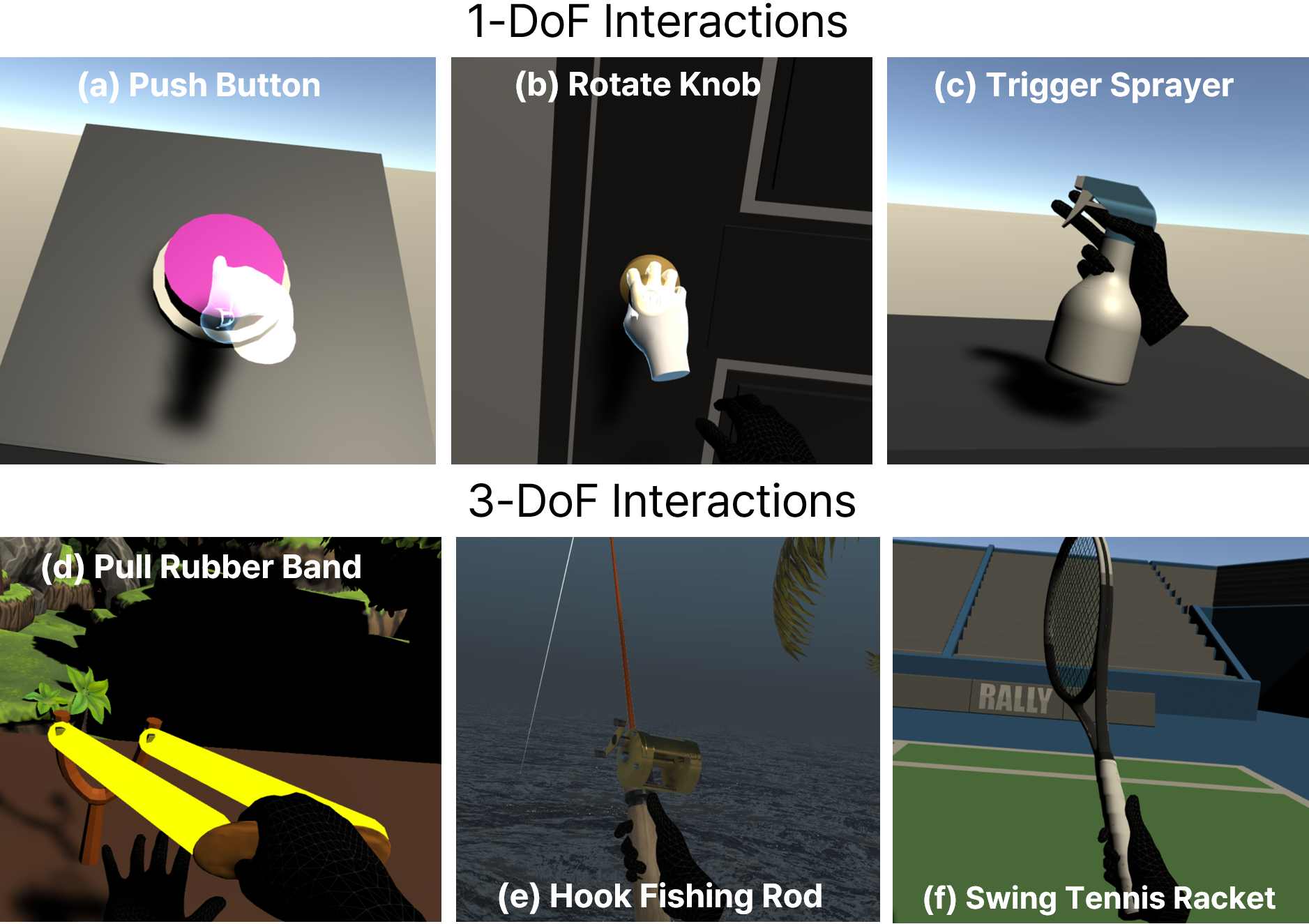}
	\caption{The six bare-hand interaction scenarios in VR. \textit{Push Button}, \textit{Rotate Knob}, and \textit{Trigger Sprayer} scenarios were designed for 1-DoF primitive hand interactions, and the \textit{Pull Rubber Band}, \textit{Hook Fishing Rod}, and \textit{Swing Tennis Racket} scenarios were designed for evaluating the QuadStretcher in more complex 3-DoF interactions.}
	\label{fig:vrInteractionScenarios}
	\Description{The screenshots of six bare-hand interaction scenarios in VR. Top left: a hand is pushing a button. Top middle: a hand is rotating a knob. Top right: a hand is triggering a sprayer. Bottom left: a hand is pulling a rubber band. Bottom middle: a hand is hooking a fishing rod. Bottom right: a hand is swinging a tennis racket.}
\end{figure*}

For a comparative evaluation between the QuadStretcher and the existing squeezing solution \cite{pezent2019tasbi, pezent2022design, pezent2022explorations}, we reproduced squeeze-feedback device, Squeezer, adopting the base design elements of Tasbi \cite{pezent2019tasbi}. 

The Squeezer is in the form of a wristwatch, as shown in Figure \ref{fig:squeezerDevice}. The band consists of six rectangular tactors (20 $\times$ 30 $\times$ 10 mm), and the tactors are interconnected with rubber strings. The tactors are flat on their inner side and have a roller (Teflon sleeve) on their outer side.  A tension string (fishing wire, diameter = 0.37 mm, max. tension = 9 kg) runs from a pulley that is driven by a geared motor (100:1 gearbox, 6V/1.6A dc motor). This assembly is located in the head module of the device. The tension string travels over the rollers of the six tactors and extends to the opposite side of the head module, where the string is tied to a steel plate. This steel plate exerts pressure on a force sensor (FSS1500, Honeywell). The string can be either pulled or released by the pulley, and a magnetic rotary encoder (AS5600, AMS) is used to monitor the rotational angle of the pulley.

An Arduino board (AVR) runs a PID control loop at 1000 Hz to position-control the motor using the feedback signal from the encoder. A command from a PC to the Arduino controls the contraction of the string, which in turn makes the tactors squeeze the forearm. The control signal for Squeezer is the amount of tension string contraction (mm). Figure \ref{fig:squeezerProfile} shows the relationship between the amount of string contraction and the squeezing tension measured by the internal force sensor (data gathered by one of the authors). The forces on each tactor may not be uniform because the cross-section of the forearm is not completely circular. If assumed circular, a simple geometric calculation yields that the force on each tactor should be approximately 0.87 times the string tension. 

Given the variability in wrist thickness among users, participants in the user study underwent an initial calibration process to determine the minimum string tension. Using two physical buttons, one for contracting and the other for releasing the string, participants manually adjusted the string's length. Once they established the minimal tension, it was set as a reference level (Figure \ref{fig:squeezerProfile}, representing a state where the string contraction is at zero). Subsequently, participants set the maximum string tension that caused no discomfort. The control signal from the PC was the amount of string contraction (mm), calculated based on the predefined min/max range of string contraction.

\section{Bare-Hand Interaction Scenarios}
\label{bareHandInteractionScenarios}

We determined bare-hand interaction scenarios for evaluating the effectiveness of QuadStretcher in comparison with Squeezer. We considered the diversity of hand motions, with the aim of encompassing a range of hand motion primitives commonly associated with AR/VR interactions.

Firstly, we included three scenarios to assess 1-DoF interactions. The chosen scenarios, \textit{Push Button}, \textit{Rotate Knob}, and \textit{Trigger Sprayer}, involve primitive hand motions~\cite{pezent2022explorations,pezent2019tasbi} such as finger pushing, wrist rotating, and finger bending that elicit a response corresponding to 1-DoF single-axis values (e.g., the amount of button pressed, the degree of knob rotated, and the extent of sprayer triggered).

Additionally, to explore more complex activities, we also designed three scenarios representing 3-DoF interactions: \textit{Pull Rubber Band}, \textit{Hook Fishing Rod}, and \textit{Swing Tennis Racket}. In these scenarios, hand actions led to 3-DoF forces that could be represented as three-dimensional vectors. The \textit{Pull Rubber Band} scenario was designed to cover position-dependent force: the force vector had the direction from the hand's position where the rubber band was pulled to the band's neutral position, with magnitude corresponding to the pulled length. The \textit{Hook Fishing Rod} scenario was designed to cover velocity-dependent force typically associated with air/motor friction~\cite{long1999velocity} for upward/downward hand movement: the force vector was oriented from the tip of the fishing rod to the underwater fish, with magnitude corresponding to the hand's velocity. In this way, the stimulation of QuadStretcher, or Squeezer, quickly reached its maximum intensity when executing a rapid hooking motion. The \textit{Swing Tennis Racket} was similarly covering the velocity-dependent force representation, but with left/right hand movement: the force vector was directed opposite to the velocity of the racket's momentum center point during forehand and backhand swings, and had a magnitude corresponding to the velocity of the hand.

\subsection{Stimuli Rendering: 1-DoF Interactions}

The 1-DoF value corresponding to the real-time hand action (i.e., the amount of button pressed, the degree of knob rotated, or the extent of sprayer triggered) ranged from 0 (neutral) to 1 (fully pressed/rotated/triggered). These values were then transformed into the control signal for the QuadStretcher or Squeezer.

For example, when a button is pressed to 40\% of its maximum, it results in a 1-DoF value of 0.4. In the QuadStretcher, the value of 0.4 represents a 40\% expansion or contraction, depending on the rendering scheme, for all four (dorsal, right, ventral, and left) stretch units, resulting in a control input of 3.44 mm (i.e., 40\% of the maximum stretch of 8.6 mm) provided to the QuadStretcher. For the Squeezer, the value was determined based on each individual's preset maximum squeezing level, which represented the maximum contraction of the tension string. For example, if someone sets their maximum contraction of the string to 10 mm, the value of 0.4 is converted to a control input of 4 mm.

\subsection{Stimuli Rendering: 3-DoF Interactions}

\begin{figure*}[t]
	\centering
	\includegraphics[width=13.5cm]{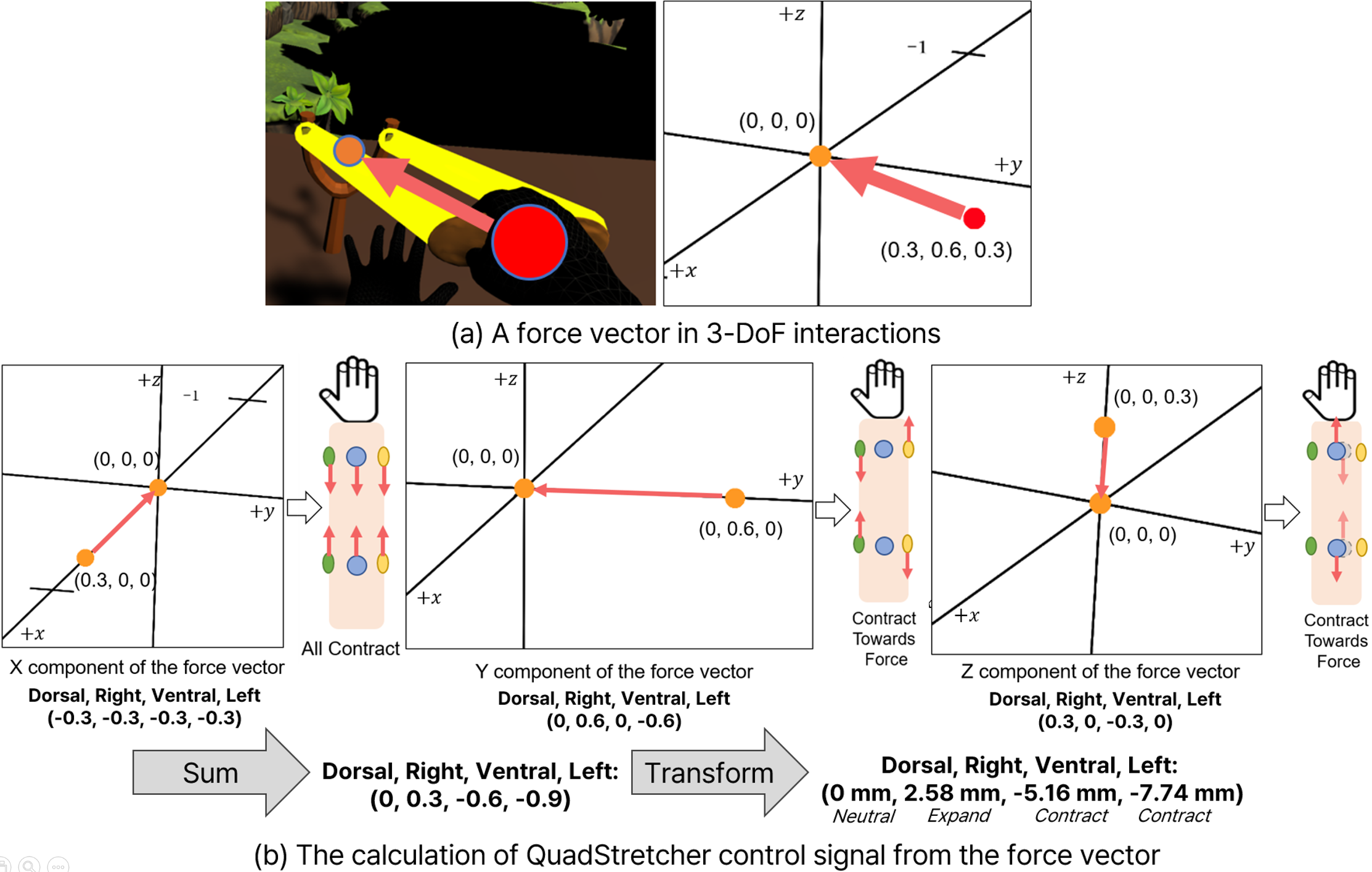}
	\caption{The stimuli rendering algorithm for QuadStretcher in 3-DoF interaction scenarios. (a) An example scenario of \textit{Pull Rubber Band} with a force vector visualized. (b) The calculation of stretch input from the force vector. The x-component of the vector is mapped to the \textit{All Contract} rendering, and the y- and z-component of the vector are mapped to the \textit{Contract Towards Force} rendering scheme. Each component of the force vector contributes to the final control signal.}
	\label{fig:renderingMethod}
	\Description{The description of how to calculate the QuadStretcher control signal from the real-time force vector. The top image shows a hand pulling a rubber band towards certain direction. The calculation of the stretch input is made from each component of the force vector.}
\end{figure*}

In 3-DoF Interaction scenarios, the 3D force vector was used to compute the control signal for the QuadStretcher or Squeezer. 

Figure \ref{fig:renderingMethod} illustrates the calculation process of the QuadStretcher's control signal from a force vector. In the case of the Squeezer, the magnitude of the force vector served as the 1-DoF value, and the control signal was computed following the same approach as in the 1-DoF interactions. For the QuadStretcher, each component of the vector played a role in different rendering layers. Firstly, the x-value was the ``non-directional" component, indicating arm movement either forward or backward in relation to the user's body. This was mapped to the \textit{All Contract} scheme, akin to the 1-DoF value. Secondly, the y-value was the ``horizontal" component, signifying arm movement to the right or left relative to the user. The \textit{Contract Towards Force} scheme was employed here, involving contraction of the stretch unit in the force direction and expansion of the stretch unit in the opposite direction. When the user pulled the rubber band to the right, the \textit{Left} stretch unit contracted while the \textit{Right} stretch unit expanded. Lastly, the z-value was the ``vertical" component, indicating arm movement upwards or downwards. When the user pulled the rubber band upwards, the \textit{Ventral} stretch unit contracted, and the \textit{Dorsal} stretch unit expanded. Each component of the force vector contributed to the calculation of the final control signal of the QuadStretcher. 

We expected the \textit{Contracts Toward Force} rendering scheme can be effective in expressing the direction of the force by employing the contraction and expansion together in an opposite way to emphasize the force direction. We compare multiple rendering options of the QuadStretcher in the following Section \ref{preliminaryStudy}.

\begin{figure}[b]
	\centering
	\includegraphics[width=8.4cm]{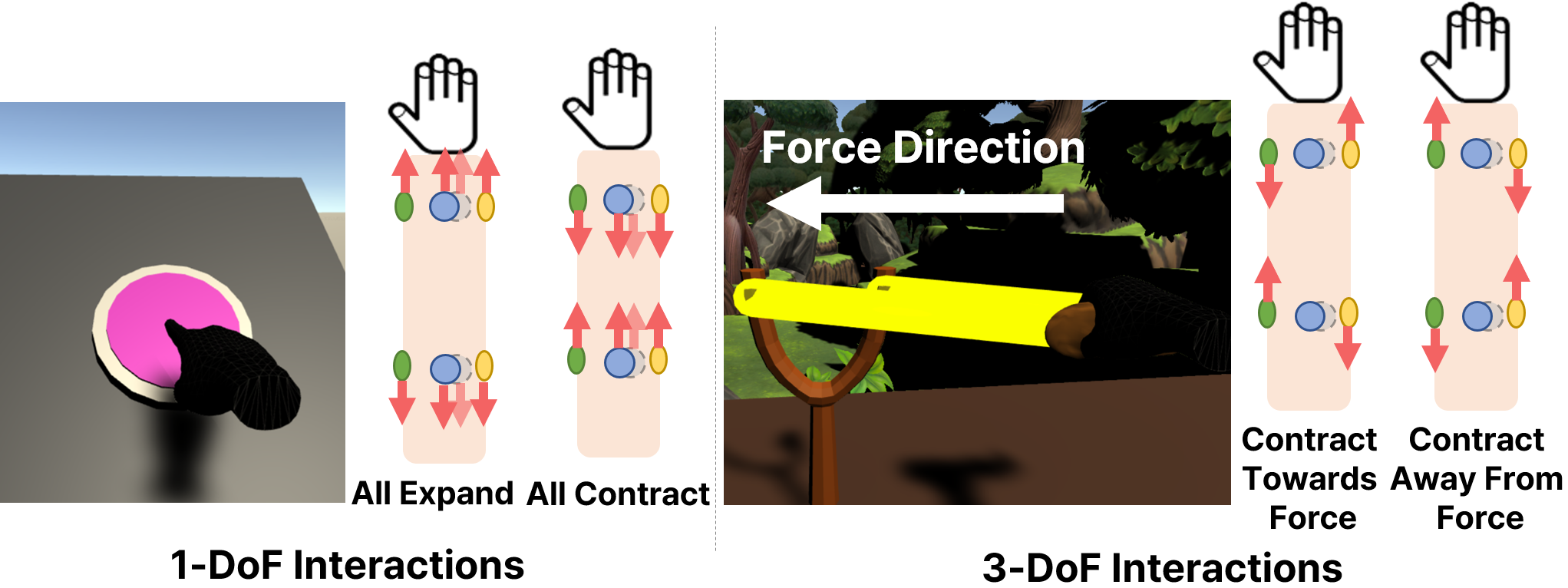}
	\caption{Tested rendering methods in the preliminary study. (a) For 1-DoF interactions, two rendering methods of \textit{All Expand} and \textit{All Contract} were tested. (b) For 3-DoF interactions, two rendering methods of \textit{Contract Towards Force} and \textit{Contract Away From Force} were tested.}
	\label{fig:renderingSchemes}
	\Description{Left image: 1-DoF interactions, a hand is pushing a button. There are illustrations of All Expand and All Contract conditions. Right image: 3-DoF interactions, a hand is pulling a rubber band towards right direction. There are illustrations of Contract Towards Force and Contract Away From Force schemes.}
\end{figure}

\subsection{Apparatus}

The pipeline of the QuadStretcher system is shown in Figure \ref{fig:quadStretchSystemPipeline}, which was the same for the Squeezer. Each bare-hand interaction scenario was implemented with Unity Engine, and a Meta Quest 2 VR headset was used to present the VR scene. During the experiment, the PC was connected to the Quest 2 through Oculus Link so that it could control the QuadStretcher and Squeezer device while running the application. We used the built-in vision-based hand tracking of the Meta Quest.

\section{Preliminary Study: Determining the Stimuli Rendering of QuadStretcher}
\label{preliminaryStudy}

\begin{figure*}[t]
	\centering
	\includegraphics[width=16cm]{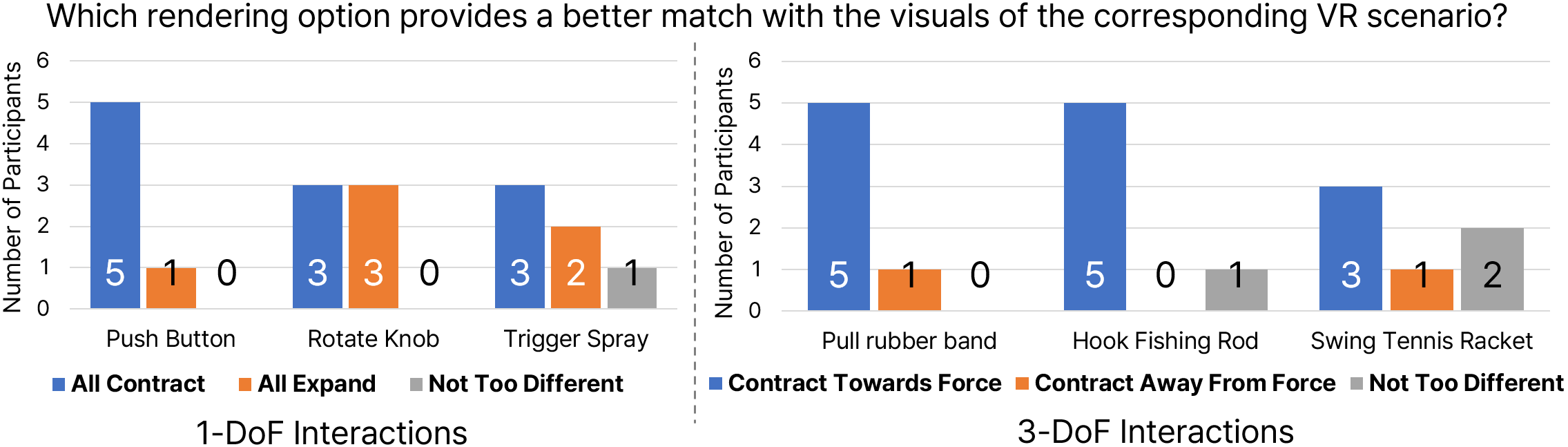}
	\caption{The result of the preliminary study.}
	\label{fig:preliminaryStudyResult}
	\Description{The bar plots of the preliminary study results. For 1-DoF interactions, (All Contract, All Expand, Not Too Different) conditions are voted for (5,1,0) for Push Button scene, (3,3,0) for Rotate Knob scene, and (3,2,1) for Trigger Spray scene. For 3-DoF interactions, (Contract Towards Force, Contract Away From Force, Not Too Different) was voted for (5,1,0) for Pull Rubber Band scene, (5,0,1) for Hook Fishing Rod scene, and (3,1,2) for Swing Tennis Racket scene.}
\end{figure*}

The purpose of this preliminary study is to determine the stimuli rendering method for the QuadStretcher. For 1-DoF interactions, two rendering options of either contracting all four stretching units (\textit{All Contract}) or expanding (\textit{All Expand}) were tested in response to the hand actions of button pressing, knob rotating, and sprayer triggering. For 3-DoF interactions, two rendering options were also tested. One was \textit{Contract Towards Force} scheme which contracts the stretching unit at the force direction and expands the stretching unit in the opposite side of the forearm. The other option was \textit{Contract Away From Force}, which is the opposite. For example, when a rubber band is pulled to the right, the force acts towards the left. In the \textit{Contract Toward Force} scheme, the \textit{Left} stretch unit contracted and the \textit{Right} stretch unit expanded, whereas \textit{Contract Away From Force} acted in the opposite way. Tested rendering options for each 1-DoF and 3-DoF interactions are illustrated in Figure \ref{fig:renderingSchemes}.

Six participants (2 females, 4 males; age: M = 30.8 years, SD = 12.6 years) from the lab took part in the study. Each participant wore the QuadStretcher and tried two rendering options as much as they wanted for each interaction scenario. They were instructed to pick one of the two rendering options with the following criteria: \textit{``Which rendering option provides a better match with the visuals of the corresponding VR scenario?''} They were allowed to switch each rendering option multiple times before finalizing their answer. The order in which the rendering option was initially presented was counterbalanced across participants.

\subsection{Result}

As shown in Figure \ref{fig:preliminaryStudyResult}, a majority of participants chose the \textit{All Contract} rendering method in the 1-DoF interaction, and the \textit{Contract Towards Force} rendering in the 3-DoF interactions.

There were notable variations in mental models among participants, as similarly noted in previous haptic feedback studies~\cite{spelmezan2009tactile, kim2021heterogeneous}. Those who chose \textit{All Contract} in the \textit{Rotate Knob} scenario explained that they anticipated needing to exert more force on their arm as the knob was rotated further, and the expression of \textit{All Contract} rendering aligned with this expectation. On the other hand, those who opted for \textit{All Expand} mentioned that they expected their skin to undergo slight twisting while turning the knob, and the \textit{All Expand} option better described this sensation. In the \textit{Trigger Sprayer} scenario, participants who chose \textit{All Contract} commented that they anticipated feeling an isometric force on their arm while triggering the spray handle, which was more matched with \textit{All Contract} method. Meanwhile, those who selected the \textit{All Expand} option focused on how it better conveyed the feeling of water spreading away from their hand, and noted that the \textit{All Expand} better described that feeling.

After considering various mental models present, we eventually opted to use the most popular choice, \textit{All Contract} as the QuadStretcher's rendering method for 1-DoF interactions, and \textit{Contract Toward Force} as the rendering method for 3-DoF interactions in the subsequent main study.

\section{Main Study: Evaluation of QuadStretcher}

\label{sec:MainStudy}

We conducted a user evaluation of QuadStretcher in comparison with the baseline solution, Squeezer, in their ability to enhance immersive VR bare-hand experiences. Each participant experienced both haptic devices across six interaction scenarios (Figure \ref{fig:vrInteractionScenarios}), and provided subjective ratings for four metrics: \textit{Realism}, \textit{Immersiveness}, \textit{Enjoyment}, and \textit{Perception of Force Direction} using a think-aloud method. We also collected qualitative user interviews. The experimental design employed a one-way within-subject approach with an independent variable being \textit{Device Type}, consisting of the following levels:

\begin{itemize}
	\item \textit{Device Type}: QuadStretcher and Squeezer
\end{itemize}

The tested order of the \textit{Device Type} conditions was counterbalanced across participants.

\subsection{Metrics}

Participants were directed to verbally assess the QuadStretcher (or the Squeezer) on the following four metrics, adapted from the presence questionnaire in virtual environments~\cite{witmer1998measuring} with a 7-point Likert scale:
\begin{itemize}
	\item \textit{Realism}: To what degree did your experiences with the current wrist-worn haptic device in the virtual environment seem consistent with your real-world experiences?
	\item \textit{Immersiveness}: How immersive was the virtual environment experience with the current wrist-worn haptic device?
	\item \textit{Enjoyment}: How enjoyable was the virtual environment experience with the current wrist-worn haptic device?
	\item \textit{Perception of Force Direction}: To what degree were you able to feel the direction of the force with the current wrist-worn haptic device?
\end{itemize}

\subsection{Participant}

A total of 20 participants (5 females, 15 males; age: M = 24.1 years, SD = 2.1 years) were recruited from a university community. Among them, 10 participants had prior experience with VR headsets but had never used them regularly, while the remaining 10 participants had never used a VR headset before. Subjects were paid approximately 23 USD for their participation.

\subsection{Procedure}

\begin{figure}[t]
	\centering
	\includegraphics[width=7.0cm]{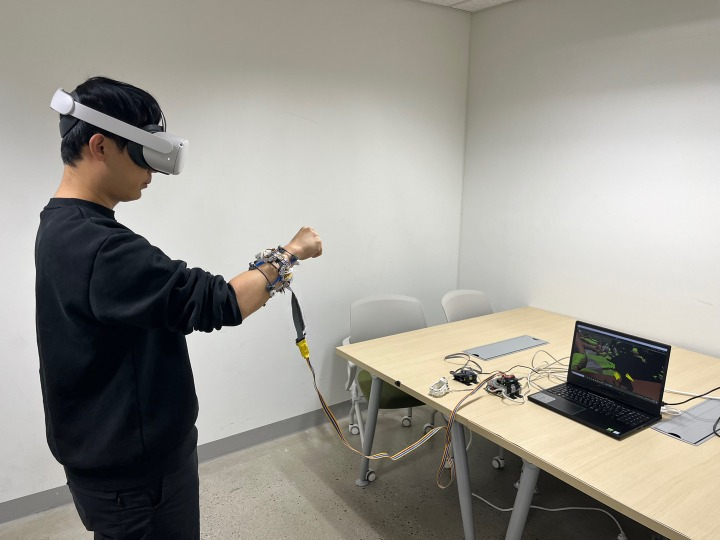}
	\caption{The user study environment.}
	\label{fig:mainStudyEnvironment}
	\Description{The environment where the user study took place. It is a silent room and a subject is wearing VR headset while wearing the QuadStretcher device as well.}
\end{figure}

After completing the demographic survey and consent form, participants were given an overview of the experiment through slides and videos. They were briefed on six different VR scenarios and were introduced to two devices. Participants were informed that they would be verbally providing responses to four metrics, \textit{Realism}, \textit{Immersiveness}, \textit{Enjoyment}, and \textit{Perception of Force Direction}, on a 7-point Likert scale. Additionally, participants were informed that the purpose of this study was to comparatively assess the experience of the two haptic devices across each VR scene.

In the initial session, participants sequentially explored all six VR scenarios without rating. Each participant wore the QuadStretcher or Squeezer, depending on the counterbalancing order, and experienced the six VR scenarios from 1-DoF interactions (\textit{Button Push}, \textit{Rotate Knob}, and \textit{Trigger Sprayer} in order) to 3-DoF interactions (\textit{Pull Rubber Band}, \textit{Hook Fishing Rod}, and \textit{Swing Tennis Racket} in order). After completing this exploration session with both the QuadStretcher and Squeezer, a 3-minute break was given. Subsequently, participants started the next session and verbally rated both the QuadStretcher and Squeezer for each VR scenario. They were allowed to explore each device as much as they wanted before providing a rating. The mean usage time for the QuadStretcher was 176 seconds, and that of Squeezer was 173 seconds for a scenario.

Participants were informed that the relative ratings between the devices were more important than absolute scores. They were informed that they could even adjust their previous rating if they wished to do so. For instance, if they initially rated the QuadStretcher as a 7 but later felt that the Squeezer was better, they could lower their rating on the QuadStretcher accordingly. After completing the ratings for each scene, an interview was conducted to gather qualitative insights into their experiences. Participants were asked about their feelings regarding the stimuli from each device, with a particular focus on identifying any notable differences between the two.

When using Squeezer, participants first set the minimum and maximum range of squeezing. Throughout the experiment, participants wore earplugs to minimize the influence of external sounds. The entire experiment took approximately 70 minutes.

\subsection{Result}

\begin{figure*}[t]
	\centering
	\includegraphics[width=11.5cm]{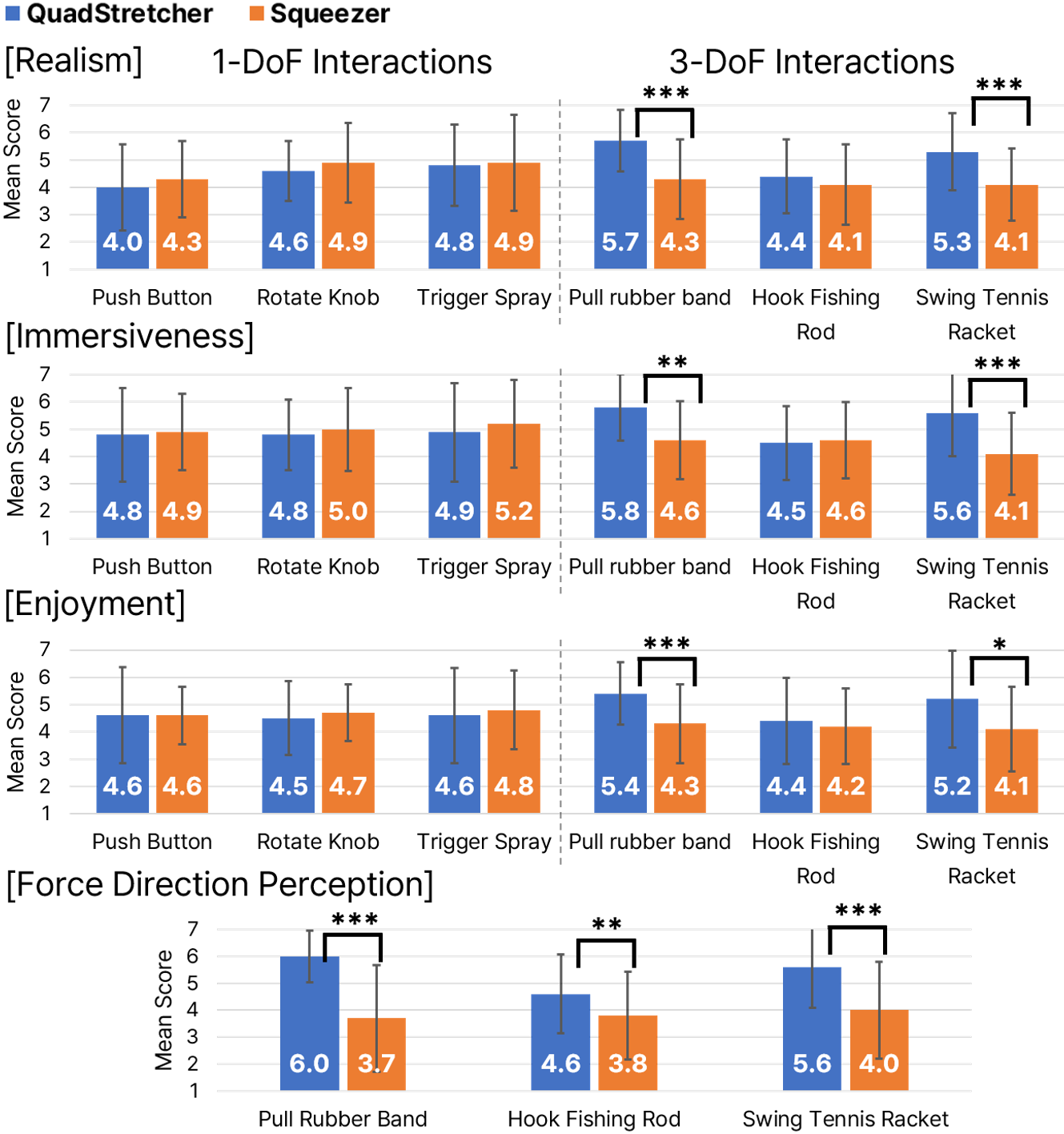}
	\caption{The subjective ratings on the four metrics collected from the main user study. The error bars show standard deviations, and asterisks $\ast$, $\ast\ast$, and $\ast\ast\ast$ indicate significant differences of \textit{p} < .05, \textit{p} < .01, and \textit{p} < .005, respectively.}
	\label{fig:mainStudyResultPlot1}
	\Description{The bar plots of the main study results. In 1-DoF interactions, there were no significant differences across the haptic devices. In 3-DoF interactions, QuadStretcher showed significantly higher ratings for Realism, Immersiveness, Enjoyment, and Force Direction Perception compared to the Squeezer.}
\end{figure*}

Figure~\ref{fig:mainStudyResultPlot1} shows bar plots representing subjective ratings on four metrics for both the QuadStretcher and Squeezer. A Friendman Test was performed to examine the statistical significance of the differences, and the corresponding statistical reports are presented in Table~\ref{table:mainStudyResult} in the Appendix section.

For 1-DoF interactions, the \textit{Realism} ratings for the QuadStretcher and Squeezer were 4.0 vs. 4.3, 4.6 vs. 4.9, and 4.8 vs. 4.9 for \textit{Push Button}, \textit{Rotate Knob}, and \textit{Trigger Spray} scenarios, respectively. The Friedman Test indicated no significant differences in subjective ratings across \textit{Device Type}, for all metrics under all VR scenarios. 

For 3-DoF interactions, the \textit{Realism} ratings for the QuadStretcher and Squeezer were 5.7 vs. 4.3, 4.4 vs. 4.1, and 5.3 vs. 4.1 for \textit{Pull Rubber Band}, \textit{Hook Fishing Rod}, and \textit{Swing Tennis Racket} scenarios, respectively. Notably, the \textit{Perception of Force Direction} ratings exhibited significant variances between the QuadStretcher and Squeezer, with ratings of 6.0 vs. 3.7, 4.6 vs. 3.8, and 5.6 vs. 4.0 for \textit{Pull Rubber Band}, \textit{Hook Fishing Rod}, and \textit{Swing Tennis Racket} scenarios, respectively. For the \textit{Pull Rubber Band} and \textit{Swing Tennis Racket} scenarios, the QuadStretcher's ratings on all metrics were significantly higher than the Squeezer, as revealed by the Friedman Test. For the \textit{Hook Fishing Rod} scenario, the QuadStretcher's rating on the \textit{Perception of the Force Direction} metric was significantly higher than the Squeezer, while ratings on other metrics were not significantly different.

\subsection{Conclusion with Qualitative Feedback}

The mean subjective ratings revealed that participants favored the QuadStretcher over the Squeezer across all metrics in complex 3-DoF hand interactions, such as \textit{Pull Rubber Band} and \textit{Swing Tennis Racket}. Meanwhile, there was no significant difference in subjective ratings for the 1-DoF uni-directional hand motions during \textit{Push Button}, \textit{Rotate Knob}, and \textit{Trigger Spray} scenarios. In the \textit{Hook Fishing Rod} scenario, the QuadStretcher received a significantly higher rating only in the \textit{Force Direction Perception} metric, while ratings on the other three metrics showed no significant difference. This result can be attributed to participants being instructed to only explore the scenario by rapidly hooking the fishing rod upward to catch a fish, leading to a hand motion closer to uni-directional rather than complex multi-axis movement.

During bare-hand VR interactions, participants were directed to articulate their experiences using the think-aloud method. Aligning with the provided ratings, most participants expressed that the QuadStretcher's haptic feedback, coordinated with dynamic 3-DoF hand motions, was highly realistic. In the \textit{Pull Rubber Band} scenario, P3 remarked, \textit{``This is very close to real. When I pull it to the right, I feel like there's a force towards the left. It's like spring. I feel the restoring force, opposite to the pulled movement.''} Similarly, P6 noted \textit{``When I changed the direction of pulling, the force I felt changed in response to that. It was much more realistic (than the Squeezer).''} Meanwhile, numerous participants stated that the Squeezer was less realistic (P1, P6, P7, P9, P10, P13, P14, P16, P20), less immersive (P5, P7), or less enjoyable (P7). P7 specifically commented that \textit{``(The Squeezer) wasn't very realistic and immersive because it applied pressure in the same manner regardless of the direction I pulled the rubber band.''} 

Similarly, in the \textit{Swing Tennis Racket} scenario, most participants favored the QuadStretcher over the Squeezer. Five participants (P1, P6, P9, P13, P20) specifically highlighted that they could feel the distinct feedback between swinging directions, particularly between forehand and backhand. P13 mentioned that, \textit{``This device (QuadStretcher) offers a distinct `pulling' direction for my arm based on the swinging direction. This makes it more realistic and enjoyable.''} In contrast, many participants (P1, P4, P6, P7, P8, P11, P13, P14, P15, P17, P18, P19) stated that the Squeezer was less realistic or immersive. Both P1 and P15 specifically noted that they could not feel the direction of the force at all with the Squeezer.

In conclusion, we could confirm that the QuadStretcher is capable of supporting immersive bare-hand experiences especially suited for interactions with higher DoF dynamics.

\section{Design Insights \& Takeaways: Forearm-Haptic Solution}

In this section, we organized design insights obtained from think-aloud interviews conducted during the main user study with 20 participants. Beyond a mere comparison to determine the superiority between the two devices, we distilled design takeaways for the future development of forearm-haptic solutions for advanced bare-hand experiences.

\textbf{(1) Skin Stretch vs. Squeeze. The inherent difference in sensation.} Aside from the degrees of freedom that each device could express, skin stretch and squeeze are inherently different haptic modalities. Consequently, participants expressed unique feelings that each type of haptic stimuli conveyed. Descriptions of the QuadStretcher's sensations include: \textit{``The QuadStretcher better conveys the feeling of the tautness of a thick, slingshot rubber band.''} by P8, \textit{``[During Push Button] This device (QuadStretcher) describes the sensation of resistance more clearly (compared to the Squeezer)''} by P7, and \textit{``The QuadStretcher’s stimuli were good for the action with movement, while Squeezer’s stimuli were good for holding things.''} by P2. Description of the Squeezer's sensation includes: \textit{``[During Push Button] I prefer this (Squeezer) because it provides a softer and more continuous feeling.''} by P6, \textit{``The Squeezer delivered a more delicate texture when water spreads out from the sprayer.''} by P8, and \textit{``Indeed, the constant pressure on my arm was not a very pleasant feeling.''} by P10. Although individual perceptions of each type of haptic stimuli varied greatly, we observed that some participants expressed that the squeeze was more suitable for a softer sensation, while the skin stretch was better suited for expressing resistance, such as in the case of a pulled/pushed object. Consequently, designers need to consider the unique sensations offered by different types of haptic stimuli, beyond the expressive capabilities associated with the device's degrees of freedom. Tailoring the haptic modality to suit a specific target application may lead to improved user experiences.

\textbf{(2) Wrist or Forearm? Where to put it on?} We must note that the two devices were worn at different locations on the forearm. While the existing solution~\cite{pezent2019tasbi} was presented as a wrist-worn device, the QuadStretcher proposed in this study was designed for the longitudinal stretching of the forearm and, as a result, is worn on the forearm lower than the wrist. Despite both being ``indirect'' forearm-haptic solutions (i.e., meaning that they provide feedback indirectly to a location other than the hands), the specific area where the haptic feedback is delivered can significantly influence user experiences. We elucidated this design factor and the corresponding user experiences through interviews. In general, participants found the QuadStretcher more natural when they thought that the interaction required force on the entire arm, such as when swinging a tennis racket (P1, P4, P8, P11, P13, P18, P20). Conversely, they considered the Squeezer more natural when they believed the action mostly required force only on the wrist, such as when rotating a knob (P1, P8, P11, P13, P15, P17, P19). P13 noted, \textit{``While I favored the Squeezer for situations that mostly involve wrist movements, such as when rotating a knob or pushing a button, I favored the QuadStretcher for activities that engage the entire arm, such as tennis or fishing.''} These observations indicate that the optimal stimulation area can differ depending on the specific bare-hand activity being targeted. Taking into account these locational aspects can be beneficial for the design of AR/VR haptic applications. 

\textbf{(3) Individual mental models vary greatly.} During the study, some participants (P1, P3, P11, P13, P17) commented that they perceived the \textit{Trigger Sprayer} scenario as necessitating the active use of arm muscles, finding the skin stretch on the forearm to be a more realistic representation. On the other hand, another group of participants (P10, P12, P15) viewed the act of triggering a sprayer as involving minimal engagement of arm muscles, believing that the simple squeeze on the wrist was more natural. This contrast highlights how the interpretation of anticipated haptic sensations in a given activity can greatly influence user satisfaction. Recognizing the diverse mental models held by individuals~\cite{spelmezan2009tactile}, providing a clear description of the haptic feedback in alignment with a particular action may unify the user interpretations, potentially fostering balanced immersion across individuals.

\section{Future Work And Limitation}
\label{futureWorkAndLimitation}

The current total range of tactor movement is 44 mm, whereas the stretching unit spans 136 mm in width. This initial design choice was made to ensure sufficient longitudinal contraction and expansion of the forearm skin, even though the length of each stretching unit can be reduced up to 85 mm based on the current mechanical structure. Minimizing the device is surely viable, but it is crucial to investigate how much the stretching unit can be shortened without losing its expressive power. This follow-up study will unveil the miniaturization limit of the QuadStretcher hardware.

Another thing to note is that the device relies on friction between the tactors and the user's forearm for grounding. The level of friction on the skin can differ among individuals due to variations in smoothness, the presence of hair, and elasticity, even under the same pressure from the rubber band. Although no noticeable instances of skin slippage were observed in our user studies, it may potentially occur for individuals with exceptionally smooth and dry skin. In such instances, it becomes crucial to exert sufficient pressure to ensure proper contact and to monitor for any discomfort related to this situation.

We also note that our exploration of the QuadStretcher's capabilities was not exhaustive, leaving more opportunities to be tested. In the current study, the expression of impact force, such as hitting a tennis ball with a racket, punching a sandbag, or firing a gun, has not been tested. These impact expressions of the QuadStretcher should be further evaluated. Additionally, the evaluation in this study focused solely on the qualitative experience of user immersion. However, exploring other aspects of the haptic device, such as its impact on quantitative task performance like pointing speed/accuracy or its efficacy in eyes-free interaction scenarios like haptic guidance, has not been undertaken. These untested opportunities remain intriguing avenues for future research.

\section{Conclusion}

In this study, we proposed a novel artifact called QuadStretcher: a skin stretch display with four independently controlled stretching units surrounding the forearm, facilitating immersive bare-hand AR/VR experiences. A user assessment demonstrated that QuadStretcher outperformed the baseline solution, Squeezer, in expressing force direction and enhancing the sense of realism during dynamic 3-DoF interactions. Furthermore, we share the design insights and takeaways distilled from the qualitative user interviews, revealing the factors that influence users' immersion levels when engaging with forearm haptic solutions. This study contributes to the evolving field of bare-hand haptic solutions, offering exciting opportunities for immersive AR/VR applications.

\begin{acks}

\end{acks}

\bibliographystyle{ACM-Reference-Format}
\bibliography{main}

\appendix
\section{appendix}

\begin{table*}[h]
	\setlength{\tabcolsep}{1pt} 
	\begin{tabular} {cccccccccc}
		\hline
		& \multicolumn{9}{c}{\multirow{2}{*}{\textbf{1-DoF Interactions}}} \\
		& & & & & & & & & \\
		\cline{2-10}
		& \multicolumn{3}{c}{\textit{Push Button}} & \multicolumn{3}{c}{\textit{Rotate Knob}} & \multicolumn{3}{c}{\textit{Trigger Spray}} \\
		\cline{2-4} \cline{5-7} \cline{8-10}
		& \multirow{2}{*}{\makecell{Quad \\Stretcher}} & \multirow{2}{*}{\makecell{Squeezer}} & \multirow{2}{*}{\makecell{Significant \\Difference}} & \multirow{2}{*}{\makecell{Quad \\Stretcher}} & \multirow{2}{*}{\makecell{Squeezer}} & \multirow{2}{*}{\makecell{Significant \\Difference}} & \multirow{2}{*}{\makecell{Quad \\Stretcher}} & \multirow{2}{*}{\makecell{Squeezer}} & \multirow{2}{*}{\makecell{Significant \\Difference}} \\ 
		& & & & & & & & & \\
		\hline
 		Realism & 4.0 (1.8) & 4.3 (1.4) & X & 4.6 (1.1) & 4.9 (1.4) & X & 4.8 (1.5) & 4.9 (1.8) & X   \\
		Immersiveness & 4.8 (1.7) & 4.9 (1.4) & X & 4.8 (1.3) & 5.0 (1.5) & X & 4.9 (1.8) & 5.2 (1.6) & X   \\
		Enjoyment & 4.6 (1.8) & 4.6 (1.1) & X & 4.5 (1.4) & 4.7 (1.0) & X & 4.6 (1.7) & 4.8 (1.4) & X   \\
		\hline
		& \multicolumn{9}{c}{\multirow{2}{*}{\textbf{3-DoF Interactions}}} \\
		& & & & & & & & & \\
		\cline{2-10}
		& \multicolumn{3}{c}{\textit{Pull Rubber Band}} & \multicolumn{3}{c}{\textit{Hook Fishing Rod}} & \multicolumn{3}{c}{\textit{Swing Tennis Racket}} \\
		\cline{2-4} \cline{5-7} \cline{8-10}
		& \multirow{2}{*}{\makecell{Quad \\Stretcher}} & \multirow{2}{*}{\makecell{Squeezer}} & \multirow{2}{*}{\makecell{Significant \\Difference}} & \multirow{2}{*}{\makecell{Quad \\Stretcher}} & \multirow{2}{*}{\makecell{Squeezer}} & \multirow{2}{*}{\makecell{Significant \\Difference}} & \multirow{2}{*}{\makecell{Quad \\Stretcher}} & \multirow{2}{*}{\makecell{Squeezer}} & \multirow{2}{*}{\makecell{Significant \\Difference}} \\ 
		& & & & & & & & & \\
		\hline
		\multirow{3}{*}{Realism} & \multirow{3}{*}{5.7 (1.1)} & \multirow{3}{*}{4.3 (1.5)} & \multirow{3}{*}{\makecell{O \\ (\textit{p} < .005, \\ $\chi^2$(1) = 8.067)}} & \multirow{3}{*}{4.4 (1.4)} & \multirow{3}{*}{4.1 (1.5)} & \multirow{3}{*}{X} & \multirow{3}{*}{5.3 (1.4)} & \multirow{3}{*}{4.1 (1.3)} & \multirow{3}{*}{\makecell{O \\ (\textit{p} < .005, \\ $\chi^2$(1) = 11.267)}} \\
		& & & & & & & & & \\
		& & & & & & & & & \\
		\hline
		\multirow{3}{*}{Immersiveness} & \multirow{3}{*}{5.8 (1.2)} & \multirow{3}{*}{4.6 (1.4)} & \multirow{3}{*}{\makecell{O \\ (\textit{p} < .01, \\ $\chi^2$(1) = 8.067)}} & \multirow{3}{*}{4.5 (1.4)} & \multirow{3}{*}{4.6 (1.4)} & \multirow{3}{*}{X} & \multirow{3}{*}{5.6 (1.6)} & \multirow{3}{*}{4.0 (1.5)} & \multirow{3}{*}{\makecell{O \\ (\textit{p} < .005, \\ $\chi^2$(1) = 11.267)}}   \\
		& & & & & & & & & \\
		& & & & & & & & & \\
		\hline
		\multirow{3}{*}{Enjoyment} & \multirow{3}{*}{5.4 (1.1)} & \multirow{3}{*}{4.3 (1.4)} & \multirow{3}{*}{\makecell{O \\ (\textit{p} < .005, \\ $\chi^2$(1) = 11.267)}} & \multirow{3}{*}{4.4 (1.6)} & \multirow{3}{*}{4.2 (1.4)} & \multirow{3}{*}{X} & \multirow{3}{*}{5.2 (1.8)} & \multirow{3}{*}{4.1 (1.5)} & \multirow{3}{*}{\makecell{O \\ (\textit{p} < .05, \\ $\chi^2$(1) = 6.250)}}   \\
		& & & & & & & & & \\
		& & & & & & & & & \\
		\hline
		\multirow{3}{*}{\makecell{Perception of \\Force \\Direction}} & \multirow{3}{*}{6.0 (1.0)} & \multirow{3}{*}{3.7 (2.0)} & \multirow{3}{*}{\makecell{O \\ (\textit{p} < .005, \\ $\chi^2$(1) = 11.842)}} & \multirow{3}{*}{4.6 (1.5)} & \multirow{3}{*}{3.8 (1.6)} & \multirow{3}{*}{\makecell{O \\ (\textit{p} < .01, \\ $\chi^2$(1) = 7.200)}} & \multirow{3}{*}{5.6 (1.5)} & \multirow{3}{*}{4.0 (1.8)} & \multirow{3}{*}{\makecell{O \\ (\textit{p} < .005, \\ $\chi^2$(1) = 9.941)}}   \\
			& & & & & & & & & \\
		& & & & & & & & & \\
		\hline
	\end{tabular}
	\caption{The result of the main user study on the comparative evaluation of the QuadStretcher and Squeezer (Section \ref{sec:MainStudy}). For the 1-DoF interactions of \textit{Push Button}, \textit{Rotate Knob}, and \textit{Trigger Spray}, the Friedman Test revealed no significant differences in subjective ratings across the QuadStretcher and Squeezer for all metrics, i.e., \textit{Realism}, \textit{Immersion}, and \textit{Enjoyment}, under all VR scenarios. However in 3-DoF interactions of \textit{Pull Rubber Band} and \textit{Swing Tennis Racket}, the Friedman Test revealed significant differences in subjective ratings across the QuadStretcher and Squeezer, for all metrics. In the \textit{Hook Fishing Rod}, the rating for the  \textit{Perception of Force Direction} metric on the QuadStretcher was significantly higher than the Squeezer. The p-values with chi-square values are listed in the rightmost column of tables for each interaction scenario.}
	\label{table:mainStudyResult}
\end{table*}

\end{document}